\documentclass[epj]{svjour}
\pdfoutput=1
\usepackage[utf8]{inputenc}
\usepackage[T1]{fontenc}
\usepackage{graphicx}
\usepackage{float}
\usepackage{amssymb}
\usepackage{amsmath}
\usepackage{hyperref}
\usepackage{color}
\definecolor{blue-violet}{rgb}{0.54, 0.17, 0.89}
\definecolor{ao(english)}{rgb}{0.0, 0.5, 0.0}
\begin{document}

\title{Wikipedia mining of hidden links between political leaders}

\author{
Klaus M. Frahm$^{1}$, Katia Jaffr\`es-Runser$^{2}$ and 
Dima L. Shepelyansky$^{1}$}

\institute{
Laboratoire de Physique Th\'eorique du CNRS, IRSAMC, 
Universit\'e de Toulouse, CNRS, UPS, 31062 Toulouse, France
\and
Institut de Recherche en Informatique de Toulouse, Universit\'e de Toulouse, INPT, Toulouse, France
}

\titlerunning{Wikipedia mining of hidden links between politicians}
\authorrunning{K.M.Frahm {\it et al.}}

\abstract{We describe a new method of reduced Google matrix
which allows to establish direct and hidden links between
a subset of nodes of a large directed network. This approach uses 
parallels with quantum scattering theory, 
developed for processes in nuclear and mesoscopic physics
and quantum chaos. The method is applied to the Wikipedia networks
in different language editions
analyzing several groups of
political leaders of USA, UK, Germany, France, Russia and G20.
We demonstrate that this approach allows to recover 
reliably direct and hidden links among political leaders.
We  argue that the reduced Google matrix method can form 
the mathematical basis for studies in social and political sciences 
analyzing  Leader-Members eXchange (LMX).
}


%\date{\today}
\date{Dated: September 7, 2016}

\maketitle

\section{Introduction}
%\label{sec1}

At present a free online encyclopaedia Wikipedia \cite{wikiorg} 
becomes the largest open source of knowledge being close to
Encyclopaedia Britanica \cite{britanica} by an accuracy of
scientific entries  \cite{giles} and overcoming the later 
by an enormous amount of available information.
A detailed analysis of strong and weak features of Wikipedia
is given at \cite{reagle,finn}.

Since Wikipedia articles make citations to each other
they generate a larger directed network
with a rather clear meaning of nodes
defined by article titles. Due to these reasons 
it is interesting to apply algorithms 
developed for search engines of World Wide Web (WWW),
those like the PageRank algorithm \cite{brin}
(see also \cite{meyer}),
to analyze the ranking properties and relations 
between Wikipedia articles. The clear meaning of 
Wikipedia nodes allows also to use its network 
as a test bed for machine learning algorithms
computing semantic relatedness \cite{gabrilovich}.

It is convenient to describe the network of $N$ Wikipedia articles 
by the Google matrix $G$ constructed from the adjacency matrix
$A_{ij}$ with elements $1$ if article (node) $j$ points to 
article (node) $i$ and zero otherwise. Then the matrix elements 
of the Google matrix take the standard form \cite{brin,meyer}
\begin{equation}
  G_{ij} = \alpha S_{ij} + (1-\alpha) / N \;\; ,
\label{eq_gmatrix} 
\end{equation}
where $S$ is the matrix of Markov transitions with elements 
$S_{ij}=A_{ij}/k_{out}(j)$, 
$k_{out}(j)=\sum_{i=1}^{N}A_{ij}\neq0$ being the node $j$ out-degree
(number of outgoing links) and with $S_{ij}=1/N$
if $j$ has no outgoing links (dangling node). 
Here $0< \alpha <1$ is the damping factor 
which for a random surfer
determines the probability $(1-\alpha)$ to
jump to any node. The properties of spectrum 
and eigenstates of $G$ have been discussed in detail
for Wikipedia and other directed networks
(see e.~g. \cite{wikispectrum,rmp2015}).

The right eigenvectors $\psi_i(j)$  of $G$ are
determined by the equation:
\begin{equation}
\label{eq_gmatrix2}
\sum_{j'} G_{jj'} \psi_i(j')=\lambda_i \psi_i(j) \; .
\end{equation}
The PageRank eigenvector $P(j)=\psi_{i=0}(j) $ corresponds to the largest 
eigenvalue $\lambda_{i=0}=1$ \cite{brin,meyer}. It has positive elements
which give the probability to find a random surfer on a given node
in the stationary long time limit of the Markov process.
All nodes can be ordered by a monotonically decreasing probability
$P(K_i)$ with the highest probability at $K=1$. The index $K$ 
is the PageRank index.  
Left eigenvectors are biorthogonal to right eigenvectors of different 
eigenvalues. 
The left eigenvector for $\lambda=1$ has identical (unit) entries 
due to the column sum normalization of $G$. 
One can show that the damping factor $\alpha$ 
in (\ref{eq_gmatrix}) only affects 
the PageRank vector (or other eigenvectors for $\lambda=1$ of $S$ in case 
of a degeneracy) while other eigenvectors are independent of $\alpha$ 
due to their orthogonality
to the left unit eigenvector for $\lambda=1$ \cite{meyer}. Thus 
all eigenvalues, except $\lambda=1$, are multiplied by a factor $\alpha$ 
when replacing $S$ by $G$. 
In the following we use the notations 
$\psi_L^T$  and $\psi_R$
for left and right eigenvectors  respectively
(here $T$ means  vector or matrix transposition).

In many real networks the number of nonzero elements
in a column of $S$ is significantly smaller then the whole matrix size $N$
that allows to find efficiently the PageRank vector by 
the PageRank algorithm of power iterations \cite{meyer}. Also
a certain number of largest eigenvalues (in modulus) and related eigenvectors
can be efficiently computed by the Arnoldi algorithm
\cite{rmp2015}.

For various language editions of
Wikipedia it was shown that the PageRank vector produces
a reliable ranking of historical figures over 35 centuries 
of $\;\;\;$  human history
\cite{wikizzs,wikievol,eomwiki9,eomwiki24} 
and  Wikipedia ranking of world universities (WRWU)
\cite{wikizzs,lageswiki}. Thus the Wikipedia ranking of historical figures
is in a good agreement with the well-known Hart ranking \cite{hart},
while the WRWU is in a good agreement with 
the Shanghai Academic ranking of world universities
\cite{shanghai}. At the same time Wikipedia ranking 
produces some new additional insight as compared to these classifications.

%%%%%%%%%%%%%%%%%%%%%%%%%%%%%%%%%%%%%%%%%%%%%%%%%%%%%
\begin{figure}
\begin{center}
\includegraphics[width=0.48\textwidth]{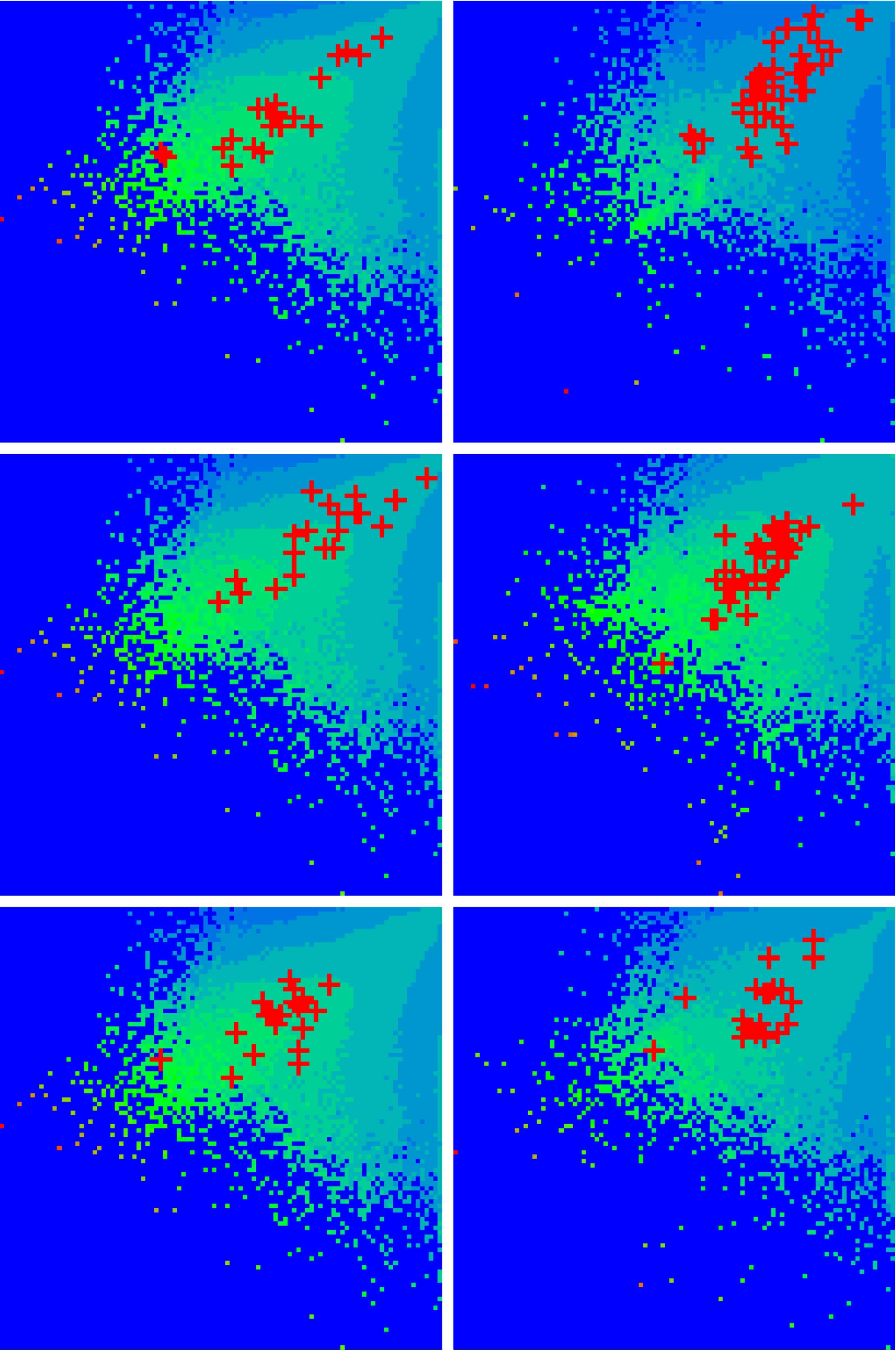}
\caption{Density of nodes  $W(K,K^*)$ on the PageRank-CheiRank plane 
averaged over $100\times100$ 
logarithmic equidistant grids for $0 \leq \ln K, \ln K^* \leq \ln N$. 
The density is averaged over all nodes inside each cell of the grid 
with the $x$-axis corresponding to $\ln K$ and the $y$-axis to 
$\ln K^*$. The panels of the left column correspond all to the 
English (Enwiki) 
and panels of the right column to the German (Dewiki, top), 
French (Frwiki, center) and Russian (Ruwiki, bottom) 
Wikipedia editions of 2013
collected in \cite{eomwiki24}. The colors represent maximum (red), 
intermediate (green) and minimum (blue) values. 
The red crosses show the positions of the selected 
reduced network nodes for 20 US (top left), 20 UK (center left), 
40 German (DE, top right), 40 French (FR, center right), 20 
Russian (RU, bottom right) politicians and 20 state leaders of G20 states 
(bottom left) with the list of names given in Tables 2-7 and 
in Figs.~2-7. }
\label{fig1}
\end{center}
\end{figure}
%%%%%%%%%%%%%%%%%%%%%%%%%%%%%%%%%%%%%%%%%%%%%%%%%%%%%

In addition to the matrix $G$ it is useful to introduce a Google matrix $G^*$
constructed from the adjacency matrix of the same network but with
inverted direction of all links. The statistical properties of the 
eigenvector $P^*$ of $G^*$ with the largest eigenvalue $\lambda=1$
have been studied first for the Linux Kernel network \cite{linux}
showing that there are nontrivial correlations between 
$P$ and $P^*$ vectors of the network.
More detailed studies have been done for Wikipedia and other networks
\cite{wikizzs,rmp2015}. The vector $P^*(K^*)$ 
is called the CheiRank vector
and the index  numbering nodes in order of monotonic
decrease of probability $P^*$ is noted as CheiRank index $K^*$.
Thus, nodes with many ingoing (or outgoing) links have small values of 
$K=1,2,3...$ (or of $K^*=1,2,3,...$) \cite{meyer,rmp2015}.
Examples of density distributions (in the $(\ln K,\ln K^*)$ plane) 
for Wikipedia editions EN, DE, FR, RU 
from the year 2013 (see network data in \cite{eomwiki24})
are shown in Fig.~\ref{fig1}.

Other eigenvectors of $G$
have $|\lambda| \leq \alpha$ \cite{meyer,rmp2015}. For $\;\;\;\;\;$ Wikipedia
is was shown that the eigenvectors with a large modulus of $\lambda$
select some specific communities of Wikipedia network 
\cite{wikispectrum,rmp2015}. However, a priory it is not possible to 
know what the meanings of these communities are.
Thus other methods are required to determine effective
interactions between $N_r$ nodes of a specific subset (group)
of the global network of a large size $N$.

Recently, the method of  reduced Google matrix
has been proposed for analysis of effective
interactions between nodes of a selected subset
embedded into a large size network \cite{greduced}.
This approach uses parallels with the quantum scattering
theory, developed for processes in nuclear and mesoscopic physics
and quantum chaos. In this work we apply this method 
to subsets (groups) of Wikipedia articles about
political leaders (politicians)
considering English, French, German and Russian 
Wikipedia editions and politicians of
USA (US), UK, Germany (DE), France (FR) and Russia (RU).
The total number of nodes for these Wikipedia networks 
is $N=4212493$ (EN), $1532978$ (DE),
$1352835$ (FR), $966284$ (RU)  \cite{eomwiki24}.
We also analyze interactions between political leaders of the G20
Los Cabos summit in 2012 \cite{g20wiki}.
The selected subsets have networks 
of $20$ or $40$ nodes and are well suited for
analysis of direct and hidden links between politicians.
In our analysis we use the Wikipedia networks collected in 2013
and described in \cite{eomwiki24}. 
The location of selected nodes on the PageRank-CheiRank
plane $(\ln K, \ln K^*)$ is shown in Fig.~\ref{fig1}.
The obtained results allow to determine interesting direct and 
hidden relations between political leaders
of the selected countries.

We should note that the analysis of interactions and relations
between political leaders represents a hot topic in social
and political sciences \cite{bass}.
Thus the interactions between leader and group members,
known as Leader-Members eXchange (LMX),
attracts at present active investigations of
researchers  in social and political sciences 
\cite{lmx1,lmx2,lmx3}. However, 
only very recently the methods
of complex networks \cite{dorogovtsev}
started to be used by in the LMX analysis
\cite{erdogan}. In this work we argue that the approach 
to determine the reduced Google matrix $G_{\rm R}$ 
represents a useful and efficient tool
for the LMX analysis of interactions inside
a group of people. Thus for a group of politicians
(a group of their articles at Wikipedia)
we find that those at the top of PageRank index $K$
are the dominant leaders being usually
country presidents or prime-ministers. It turns out 
that the obtained $G_{\rm R}$ matrix, describing the interactions 
between group members, is composed of three matrix components. 
These components 
describe: the direct interactions $G_{rr}$ between group members,
a projector part $G_{pr}$ which is mainly
imposed by the PageRank of group members
given by the global $G$ matrix
and a component $G_{qr}$ from hidden interactions between members
which appear due to indirect links via the global network.
Thus the reduced  matrix $G_{\rm R} =  G_{rr} + G_{\rm pr} + G_{\rm qr}$
allows to obtain precise information
about the group members taking into account their
environment given by the global Wikipedia network.
We think that this $G_{\rm R}$ matrix approach
provides mathematical grounds for 
the LMX studies. 

The paper is composed as follows:
Section 2 shortly describes the method of reduced Google matrix,
Section 3 presents distributions of selected
subsets on the PageRank-CheiRank plane of global and reduced networks,
Sections 4, 5, 6, 7, 8, and 9 describe the results of the 
reduced Google matrix  
analysis for politicians of USA, UK, Germany, France, Russia and G20
respectively. Section 10 provides a particular analysis for the group 
of French politicians in terms of effective networks of strongest friends 
or followers using either the matrix $G_{\rm R}$ or the hidden interactions 
given by the component $G_{\rm qr}$. 
The discussion of the obtained results is given in Section 11.
All numerical data of the reduced Google matrix of groups
of political leaders considered here are publicly available at
the web site \cite{ourwebpage}. 

\section{Reduced Google matrix}

The concept of reduced Google matrix $G_{\rm R}$
was introduced in \cite{greduced} on the basis of
the following observation. At present directed networks of 
real systems can be very large
(about  $4.2$ million articles for the English Wikipedia edition in 2013 
\cite{rmp2015} or
$3.5$ billion web pages for a publicly accessible web
crawl that was gathered by the Common Crawl Foundation in 2012
\cite{vigna}). In certain cases one may be interested in the 
particular interactions among a small reduced subset of $N_r$ nodes 
with $N_r \ll N$ instead of the interactions in the entire network. 
However, the interactions between these $N_r$ nodes should be 
correctly determined taking into account that there are many 
indirect links between the $N_r$ nodes via all other 
$N_s=N-N_r$ nodes of the network.
This leads to the problem of the reduced Google matrix $G_{\rm R}$
with $N_r$ nodes which describes the interactions of
a subset of $N_r$ nodes.

In a certain sense we can trace parallels with 
the problem of quantum scattering
appearing in nuclear and mesoscopic physics 
\cite{sokolov1989,sokolov1992,beenakker,guhr,jalabert}
and quantum chaotic scattering \cite{gaspard}. 
Indeed, in the scattering problem there are effective
interactions between open channels to localized basis states in a well 
confined scattering domain where a particle can spend a certain time 
before it escapes by open channels.
Having this analogy in mind we construct the reduced
Google matrix $G_{\rm R}$ which describes interactions 
between selected $N_r$ nodes
and satisfies the standard requirements of the Google matrix.

Let $G$ be a typical Google matrix of Perron-Frobenius type for 
a network with $N$ nodes such that $G_{ij}\ge 0$ and the 
column sum normalization $\sum_{i=1}^N G_{ij}=1$ are verified. 
We consider a sub-network 
with $N_r<N$ nodes, called ``reduced network''. In this case we can write 
$G$ in a block form~:
\begin{equation}
\label{eq_Gblock}
G=\left(\begin{array}{cc}
G_{rr} & G_{rs} \\
G_{sr} & G_{ss} \\
\end{array}\right)
\end{equation}
where the index ``$r$'' refers to the nodes of the reduced network and 
``$s$'' to the other $N_s=N-N_r$ nodes which form a complementary 
network which we will call ``scattering network''. 

We denote the PageRank vector of the full network as
\begin{equation}
\label{eq_Pageank0}
P=\left(\begin{array}{c}
P_r  \\
P_s  \\
\end{array}\right)
\end{equation}
which satisfies the equation $G\,P=P$ 
or in other words $P$ is the right eigenvector of $G$ for the 
unit eigenvalue. This eigenvalue equation reads in block notations:
\begin{eqnarray}
\label{eq_Pagerank1}
({\bf 1}-G_{rr})\,P_r-G_{rs}\,P_s&=&0,\\
\label{eq_Pagerank2}
-G_{sr}\,P_r+({\bf 1}-G_{ss})\,P_s&=&0.
\end{eqnarray}
Here ${\bf 1}$ is the unit matrix of corresponding size $N_r$
or $N_s$.
Assuming that the matrix ${\bf 1}-G_{ss}$ is not singular, i.e. all 
eigenvalues $G_{ss}$ are strictly smaller than unity (in modulus), we obtain 
from (\ref{eq_Pagerank2}) that 
\begin{equation}
\label{eq_Ps}
P_s=({\bf 1}-G_{ss})^{-1} G_{sr}\,P_r
\end{equation}
which gives together with (\ref{eq_Pagerank1}):
\begin{equation}
\label{eq_Geff1}
G_{\rm R}P_r=P_r\quad,\quad
G_{\rm R}=G_{rr}+G_{rs}({\bf 1}-G_{ss})^{-1} G_{sr}
\end{equation}
where the matrix $G_{\rm R}$ of size $N_r\times N_r$, defined for the 
reduced network, can be viewed as an effective reduced Google matrix. 
Here the contribution of $G_{rr}$ accounts for direct links 
in the reduced network and the second term with the matrix inverse 
corresponds to all contributions of indirect links of arbitrary order. 
We note that in mesocopic scattering problems one typically uses 
an expression of the scattering matrix which has a similar structure 
where the scattering channels correspond to the reduced network and 
the states inside the scattering domain to the scattering network 
\cite{beenakker}. 

The matrix elements of $G_{\rm R}$ are non-negative since the matrix 
inverse in (\ref{eq_Geff1}) can be expanded as:
\begin{equation}
\label{eq_inverse_expand}
({\bf 1}-G_{ss})^{-1}=\sum_{l=0}^\infty G_{ss}^{\,l} \;\; .
\end{equation}
In (\ref{eq_inverse_expand}) 
the integer $l$ represents the order of indirect links, i.~e. the number 
of indirect links which are used to connect indirectly two nodes of the 
reduced network. The matrix inverse corresponds to an exact resummation 
of all orders of indirect links. 
According to (\ref{eq_inverse_expand}) 
the matrix  $({\bf 1}-G_{ss})^{-1}$ and therefore also $G_{\rm R}$
have non-negative matrix elements. 
It remains to show that $G_{\rm R}$ also fulfills the condition 
of column sum  normalization being unity. 
For this let us denote by $E^T=(1,\,\ldots,\,1)$ 
the line vector of size $N$ with unit entries and by $E_r^T$ (or $E_s^T$) 
the corresponding vectors for the reduced (or scattering) network 
with $N_r$ (or $N_s$) unit entries such that $E^T=(E_r^T,\,E_s^T)$. 
The column sum normalization for the full Google matrix $G$ implies that 
$E^T G=E^T$, i.~e. $E^T$ is the left eigenvector of $G$ with eigenvalue 
$1$. This equation becomes in block notation:
\begin{eqnarray}
\label{eq_leftE1}
E_r^T({\bf 1}-G_{rr})-E_s^T G_{sr}&=&0,\\
\label{eq_leftE2}
-E_r^T G_{rs}+E_s^T({\bf 1}- G_{ss})&=&0.
\end{eqnarray}
From (\ref{eq_leftE2}) we find that $E_s^T=E_r^T G_{rs}({\bf 1}- G_{ss})^{-1}$
which implies together with (\ref{eq_leftE1}) that 
$E_r^T G_{\rm R}=E_r^{T}$ using $G_{\rm R}$ as 
in (\ref{eq_Geff1}). This shows that the column sum normalization condition is 
indeed verified for $G_{\rm R}$ justifying that this matrix is indeed 
an effective Google matrix for the reduced network. 

We can question how to evaluate practically the expression 
(\ref{eq_Geff1}) of $G_{\rm R}$ for a particular sparse and quite large 
network with a typical situation
when $N_r\sim 10^2$-$10^3$ is small compared to $N$ and $N_s \approx N\gg N_r$. 
If $N_s$ is too large (e.~g. $N_s > 10^5$) a direct naive evaluation 
of the matrix inverse $({\bf 1}- G_{ss})^{-1}$ in (\ref{eq_Geff1}) 
by Gauss algorithm is not possible. In this case we can try the 
expansion (\ref{eq_inverse_expand}) provided it converges sufficiently 
fast with a modest number of terms. However, this is most likely not the 
case for typical applications since $G_{ss}$ is likely to have 
at least one eigenvalue very close to unity. 

Therefore, we consider the situation 
where the full Google matrix has a well defined gap between the leading 
unit eigenvalue and the second largest eigenvalue (in modulus). For example 
if $G$ is defined using a damping factor $\alpha$ in the standard way, 
as in (\ref{eq_gmatrix}), the 
gap is at least $1-\alpha$ which is $0.15$ for the standard choice 
$\alpha=0.85$ \cite{meyer}. 
For such a situation we expect that the matrix $G_{ss}$ 
has a leading real eigenvalue close to unity 
(but still different from unity so
that ${\bf 1}-G_{ss}$ is not singular) 
while the other eigenvalues are clearly 
below this leading eigenvalue with a gap comparable to the gap of 
the full Google matrix $G$. In order to evaluate the expansion 
(\ref{eq_inverse_expand}) efficiently, we need to take out analytically 
the contribution of the leading eigenvalue close to unity which is 
responsible for the slow convergence. 

%%%%%%%%%%%%%%%%%%%%%%%%%%%%%%%%%%%%%%%%%%%%%%%%%%%%%
\begin{table}
\begin{center}
\caption{Values of $1-\lambda_c$ and of $\Sigma_P$ for the 
five groups of politicians and four cases of G20 subnetworks with 
$\lambda_c$ being the leading eigenvalue of the matrix $G_{ss}$ and 
$\Sigma_P=\Vert P_r\Vert_1$ being the 1-norm of the projected PageRank 
representing the relative PageRank weight of the subset in the 
full network. }
\label{tab1}
\begin{tabular}{|l|l|c|c|}
\hline
Network & Group & $1-\lambda_c$ & $\Sigma_P$ \\
\hline
\hline
Enwiki & Politicians US & 0.0003680 & 0.0003787 \\
\hline
Enwiki & Politicians UK & 0.0001041 & 0.0001068 \\
\hline
Dewiki & Politicians DE & 0.0002705 & 0.0002802 \\
\hline
Frwiki & Politicians FR & 0.0003810 & 0.0004005 \\
\hline
Ruwiki & Politicians RU & 0.0003136 & 0.0003247 \\
\hline
Enwiki & G20 EN & 0.0002465 & 0.0002508 \\
\hline
Dewiki & G20 DE & 0.0001492 & 0.0001513 \\
\hline
Frwiki & G20 FR & 0.0002131 & 0.0002159 \\
\hline
Ruwiki & G20 RU & 0.0002707 & 0.0002734 \\
\hline
\end{tabular}

\end{center}
\end{table}
%%%%%%%%%%%%%%%%%%%%%%%%%%%%%%%%%%%%%%%%%%%%%%%%%%%%%

Below we denote by $\lambda_c$ this leading eigenvalue and by $\psi_R$ 
($\psi_L^T$) the corresponding right (left) eigenvector such that 
$G_{ss}\psi_R=\lambda_c\psi_R$ (or $\psi_L^T G_{ss}=\lambda_c\psi_L^T$). 
Both left and right eigenvectors as well as $\lambda_c$ can be efficiently 
computed by the power iteration method in a similar way as the standard 
PageRank method. We note that one can easily  show that 
$\lambda_c$ must be real and that both left/right eigenvectors can be chosen 
with positive elements. Concerning the normalization for $\psi_R$ we 
choose $E_s^T\psi_R=1$ and for $\psi_L$ we choose $\psi_L^T\psi_R=1$. 
It is well known (and easy to show) that $\psi_L^T$ is orthogonal to all other 
right eigenvectors (and $\psi_R$ is orthogonal to all other 
left eigenvectors) of $G_{ss}$ with eigenvalues different from $\lambda_c$. 
We introduce the operator ${\cal P}_c=\psi_R\psi_L^T$ which is the 
projector onto the eigenspace of $\lambda_c$ and we denote by 
${\cal Q}_c={\bf 1}-{\cal P}_c$ the complementary projector. 
One verifies directly that both projectors commute with the matrix $G_{ss}$ 
and in particular ${\cal P}_c G_{ss}=G_{ss}{\cal P}_c=\lambda_c{\cal P}_c$. 
Therefore we can write:
\begin{eqnarray}
\label{eq_inverse_project1}
({\bf 1}-G_{ss})^{-1}&=&({\cal P}_c+{\cal Q}_c)({\bf 1}-G_{ss})^{-1}
({\cal P}_c+{\cal Q}_c)\\
\label{eq_inverse_project2}
&=&{\cal P}_c\frac{1}{1-\lambda_c}+
{\cal Q}_c({\bf 1}-G_{ss})^{-1}{\cal Q}_c\\
\label{eq_inverse_project3}
&=&{\cal P}_c\frac{1}{1-\lambda_c}+
({\bf 1}-\bar G_{ss})^{-1}{\cal Q}_c\\
\label{eq_inverse_project4}
&=&{\cal P}_c\frac{1}{1-\lambda_c}+
{\cal Q}_c \sum_{l=0}^\infty \bar G_{ss}^{\,l}
\end{eqnarray}
with $\bar G_{ss}={\cal Q}_c G_{ss}{\cal Q}_c$ and using the 
standard identity ${\cal P}_c{\cal Q}_c=0$ for complementary 
projectors. 
The expansion in (\ref{eq_inverse_project4}) has the advantage that 
it converges rapidly since $\bar G_{ss}^{\,l}\sim |\lambda_{c,2}|^l$ 
with $\lambda_{c,2}$ being the second largest eigenvalue which is 
significantly 
lower than unity (e.~g. $|\lambda_{c,2}|\approx \alpha=0.85$ for the case 
with a damping factor). The first contribution due to the leading 
eigenvalue $\lambda_c$ close to unity is taken out analytically once 
the left and right eigenvectors, and therefore also the projector 
${\cal P}_c$, are known. 

The combination of (\ref{eq_Geff1}) and (\ref{eq_inverse_project4}) 
provides an explicit algorithm feasible for a numerical implementation 
for the case of modest values of $N_r$, large values of $N_s$ and of course 
for sparse matrices $G$, $G_{ss}$, etc.  The described method  
can also be modified to take out analytically the contributions 
of several leading eigenvalues close to unity as described in \cite{greduced}.
We note that the numerical methods described in \cite{fgsjphysa}
allow to determine the eigenvalues $\lambda_c$ 
(and corresponding eigenvectors) 
which are exponentially close to unity
(e.~g. $1-\lambda_c \sim 10^{-16}$) so that the expression 
(\ref{eq_inverse_project1}) can be efficiently computed numerically.

On the basis of the above equations 
(\ref{eq_Geff1})-(\ref{eq_inverse_project1})
the reduced Google matrix can be presented as a sum of three components
\begin{equation}
\label{eq_3terms}
G_{\rm R}=G_{rr} + G_{\rm pr} + G_{\rm qr} ,
\end{equation}
with the first component $G_{rr}$ given by direct matrix elements of $G$
among the selected $N_r$ nodes.
The second projector component $G_{\rm pr}$ is given by
\begin{equation}
\label{eq_2ndterm}
G_{\rm pr} =  G_{rs}  {\cal P}_c G_{sr}/(1-\lambda_c) , \; 
{\cal P}_c=\psi_R\psi_L^T \;.
\end{equation}
We mention that this contribution is of the form 
$G_{\rm pr}=\tilde\psi_R\tilde\psi_L^T/(1-\lambda_c)$ 
with $\tilde\psi_R=G_{rs}\,\psi_R$ and 
$\tilde\psi_L^T=\psi_L^TG_{sr}$ being two small vectors 
defined on the reduced space of dimension $N_r$. Therefore 
$G_{\rm pr}$ is indeed a (small) matrix of rank one which is also 
confirmed by a numerical diagonalization of this matrix. 
The third component $G_{\rm qr}$ of indirect or hidden links is given by
\begin{equation}
\label{eq_3rdterm}
G_{\rm qr} =  G_{rs} [{\cal Q}_c \sum_{l=0}^\infty \bar G_{ss}^{\,l}]  G_{sr} , \; 
{\cal Q}_c={\bf 1}-{\cal P}_c, \;
\bar G_{ss}={\cal Q}_c G_{ss}{\cal Q}_c .
\end{equation}

Even though the decomposition (\ref{eq_3terms}) is at first motivated 
by the numerical efficiency to evaluate the matrix inverse, it is equally 
important concerning the interpretation of the different terms 
and especially the last contribution 
(\ref{eq_3rdterm}) which is typically rather small as compared to 
(\ref{eq_2ndterm}) but plays in an important role as we will see below. 

Concerning the numerical algorithm to evaluate all contributions 
in (\ref{eq_3terms}), we mention that we first determine 
by the power iteration method the leading left $\psi_L$ and right eigenvector 
$\psi_R$ of the matrix $G_{ss}$ which also provides an accurate value 
of the corresponding eigenvalue $\lambda_c$ or better of $1-\lambda_c$ 
(by taking the norm of the projection of $G\psi_R$ on the reduced space which 
is highly accurate even for $\lambda_c$ close to $1$). 
These two vectors provide directly $G_{\rm pr}$ by (\ref{eq_2ndterm}) 
and allow to numerically apply the projector ${\cal Q}_c$ to an 
arbitrary vector (with $\sim N$ operations). 
The most expensive part is the evaluation of the last 
contribution according to (\ref{eq_3rdterm}). For this we apply successively 
$\bar G_{ss}={\cal Q}_c G_{ss}{\cal Q}_c$ 
to an arbitrary column of $G_{sr}$ which can 
be done by a sparse matrix vector multiplication or the efficient application 
of the projector.  

Therefore, we can calculate in parallel, for each column $j$ of $G_{sr}$, 
the following product ${\cal Q}_c \sum_{l=0}^\infty \bar G_{ss}^{\,l} G_{sj}$. 
This computation can be performed using the power iterations 
algorithm of PageRank 
which converges after about \mbox{$\sim 200$} $-250$ terms. 
Indeed, the contribution of the leading eigenvalue (of $G_{ss}$) 
has been taken out and the eigenvalues of $\bar G_{ss}$ are roughly below 
the damping factor $\alpha=0.85$. In the end the resulting 
vector is multiplied 
with the matrix $G_{rs}$ which provides one column of $G_{\rm qr}$. 
This procedure has to be repeated for each of the $N_r$ columns but 
the number $N_r$ is typically very modest (20 or 40 in this work) 
and the computation of the different columns can actually be done in 
parallel on typical multicore machines. 

Concerning the choice of the reduced space we use 5 groups 
of 20 or 40 political leaders of 5 countries (US, UK, DE, FR, RU) 
for 4 Wikipedia editions 
(EN, DE, FR, RU with EN-Wikipedia for both US and UK politicians). 
We also consider the group of G20 state leaders for which we use 
all 4 of these Wikipedia editions even though here we concentrate on 
the G20 data obtained for EN-Wikipedia. 
A detailed description of these subsets is given in Section 3. 
For the data sets of politicians considered in this work we find 
that typically $1-\lambda_c\sim 10^{-4}$ and the right eigenvector 
$\psi_R$ of $G_{ss}$ is rather close to the full PageRank of $G$ 
(for the leading nodes in the full PageRank not belonging to the 
reduced space).  
Furthermore, we find that an approximate relation 
holds: $1-\lambda_c \approx \Sigma_P=\Vert P_r\Vert_1$
where $ \Sigma_P$ is the PageRank probability
of the global network concentrated on the subset of
$N_r$ selected nodes. The data of Table~\ref{tab1}
show that this relation works with an accuracy of a couple
of percent. To understand this result mathematically, we 
replace in (\ref{eq_Ps}) the matrix inverse by the first term 
of (\ref{eq_inverse_project4}) which gives 
$P_s\approx \psi_R\psi_L^T\,G_{sr}\,P_r/(1-\lambda_c)$ 
(for $N_s\gg N_r$ and $1-\lambda_c\ll 1$). 
Furthermore we note that $\psi_L\approx E_s$ 
(for most nodes) and $E_s^T\,G_{sr}\,P_r\approx E_r^T P_r=\Vert P_r\Vert_1$ 
such $P_s\approx \psi_R \Vert P_r\Vert_1/(1-\lambda_c)$. 
Since $\Vert P_s\Vert_1\approx \Vert \psi_R\Vert_1=1$ we find indeed that 
$\Vert P_r\Vert_1\approx 1-\lambda_c$. 

The numerical computations show that the vectors $\tilde\psi_R$ and 
$\tilde\psi_L^T$ introduced  below equation (\ref{eq_2ndterm}) 
are approximately given by $\tilde\psi_R=P_r$ and 
$\tilde \psi_L^T=E_r^T$ such that 
$G_{\rm pr}\approx P_r\,E_r^T/(1-\lambda_c)$ is rather 
close to a rank one Google matrix (since $\Vert P_r\Vert_1\approx 1-\lambda_c$)
and with identical columns given by the normalized vector $P_r/(1-\lambda_c)$. 
More precisely, we will indeed see in Sections 4-9, that the overall 
column sums of $G_{\rm pr}$ account for $\sim 95$-97\% of the total column 
sum of $G_{\rm R}$. In other words, in terms of probability the contribution 
of $G_{\rm pr}$ is dominant in $G_{\rm R}$ but it is also kind of trivial with 
nearly identical columns. 
Therefore the two small contributions of $G_{rr}$ and $G_{\rm qr}$ 
are indeed very important 
for the interpretation even though they only contribute weakly to the 
overall column sum normalization. 

%%%%%%%%%%%%%%%%%%%%%%%%%%%%%%%%%%%%%%%%%%%%%%%%%%%%%
\begin{figure}
\begin{center}
\includegraphics[width=0.48\textwidth]{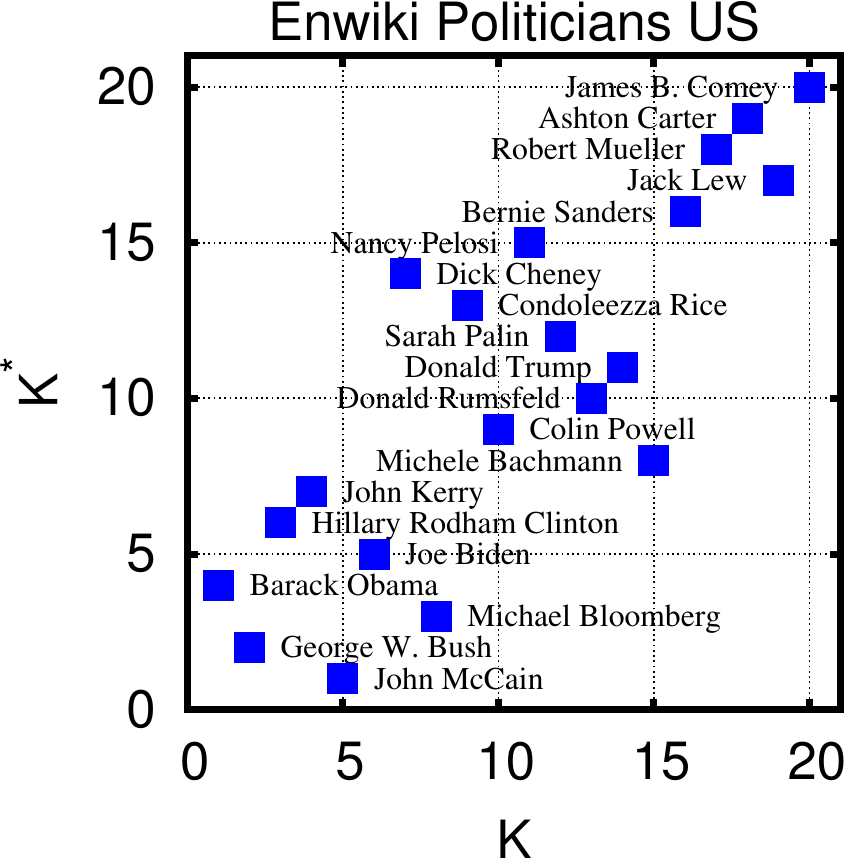}
\caption{Position of nodes in the local $(K,K^*)$ plane of the 
reduced network for 20 US politicians in the Enwiki network.
The names are shown on the same lines of the corresponding data points.
}
\label{fig2}
\end{center}
\end{figure}
%%%%%%%%%%%%%%%%%%%%%%%%%%%%%%%%%%%%%%%%%%%%%%%%%%%%%

%%%%%%%%%%%%%%%%%%%%%%%%%%%%%%%%%%%%%%%%%%%%%%%%%%%%%
\begin{table}
\begin{center}
\begin{tabular}{|l|r|r|r|}
\hline
Names (US) & $K$ & $K^*$ & $K_G$ \\
\hline
\hline
Barack Obama & 1 & 4 & 2 \\
\hline
George W. Bush & 2 & 2 & 1 \\
\hline
Hillary Rodham Clinton & 3 & 6 & 6 \\
\hline
John Kerry & 4 & 7 & 4 \\
\hline
John McCain & 5 & 1 & 3 \\
\hline
Joe Biden & 6 & 5 & 7 \\
\hline
Dick Cheney & 7 & 14 & 5 \\
\hline
Michael Bloomberg & 8 & 3 & 14 \\
\hline
Condoleezza Rice & 9 & 13 & 10 \\
\hline
Colin Powell & 10 & 9 & 12 \\
\hline
Nancy Pelosi & 11 & 15 & 8 \\
\hline
Sarah Palin & 12 & 12 & 9 \\
\hline
Donald Rumsfeld & 13 & 10 & 11 \\
\hline
Donald Trump & 14 & 11 & 20 \\
\hline
Michele Bachmann & 15 & 8 & 16 \\
\hline
Bernie Sanders & 16 & 16 & 15 \\
\hline
Robert Mueller & 17 & 18 & 13 \\
\hline
Ashton Carter & 18 & 19 & 19 \\
\hline
Jack Lew & 19 & 17 & 18 \\
\hline
James B. Comey & 20 & 20 & 17 \\
\hline
\end{tabular}

\caption{List of names of 20 selected US politicians and the 
corresponding PageRank index values $K$, $K^*$ and $K_G$ 
of the reduced network matrices $G_{\rm R}$, $G_{\rm R}^*$ and $G_{rr+qr}$ given as 
the sum of $G_{rr}$ and $G_{qr}$ (without diagonal elements). 
All matrices were computed for the English Wikipedia edition of 2013.}
\label{tab2}
\end{center}
\end{table}
%%%%%%%%%%%%%%%%%%%%%%%%%%%%%%%%%%%%%%%%%%%%%%%%%%%%%

The meaning of $G_{rr}$ is rather clear since is gives direct links
between the selected nodes. In contrast, the meaning of
 $G_{\rm qr}$ is significantly  more interesting since it generates
indirect links between the $N_r$ nodes due to their interactions with the
global network environment. We note that $G_{\rm qr}$
is composed of two parts $G_{\rm qr} = G_{\rm qrd} + G_{\rm qrnd}$
where the first diagonal term $G_{\rm qrd}$ represents
 a probability to stay on the same node
during multiple iterations of $\bar G_{ss}$ in (\ref{eq_3rdterm})
while the second nondiagonal term $G_{\rm qrnd}$ represents 
indirect (hidden) links between the $N_r$ nodes 
appearing via the global network.
We note that in principle certain matrix elements of
$G_{\rm qr}$ can be negative, which is possible due to
negative terms in ${\cal Q}_c={\bf 1}-{\cal P}_c$ appearing in
(\ref{eq_3rdterm}). However, for all subsets considered in this work
the total weight of negative elements was negligibly
small (about $10^{-10}$ for the data of UK politicians, $0$ for 
data of politicians of other countries, 
and $10^{-5}$ for the G20 state leader data, 
of the total weight $1$ for $G_{\rm R}$). 

It is convenient to characterize the strength of 3 components
in (\ref{eq_3terms}) by their respective weights 
$W_{rr}$, $W_{\rm pr}$, $W_{\rm qr}$ given respectively
by the sum of all matrix elements of   $G_{rr}$, $G_{\rm pr}$, $G_{\rm qr}$
divided by $N_r$. By definition we have $W_{rr} + W_{\rm pr} + W_{\rm qr} =1$.

In the following sections we  will see that all three
components of (\ref{eq_3terms}) play  important roles.
We present here only results for $G_{\rm R}$ obtained from $G$.
The results for the network with inverted direction of links,
corresponding to CheiRank of $G_{\rm R}^*$ of $G^*$
are given at \cite{ourwebpage}.

\section{PageRank-CheiRank plane of Wikipedia editions}

For our studies we choose 6 independent groups of articles of
20 US and 20 UK politicians from 
Enwiki, 40 German and 40 French politicians from 
Dewiki and Frwiki respectively, 20 Russian politicians from
Ruwiki and the 20 G20 state leaders from Enwiki. 
The information about number of nodes and links
for each Wikipedia edition is available at \cite{eomwiki24}.
In the selection of names of political leaders
of each country we used the names appearing at the top of Google search
on e.~g. ``politicians of Russia'', in addition
we take only those politician who were active
in the period not more than 10 - 20 years
before the collection date of our Wikipedia editions of 2013.
A few names are used to have a group of 20 or 40.
We do not pretend that we selected all important politicians
of a given country but we suppose that the main part of them
is present in our selection.

%%%%%%%%%%%%%%%%%%%%%%%%%%%%%%%%%%%%%%%%%%%%%%%%%%%%%
\begin{figure}
\begin{center}
\includegraphics[width=0.48\textwidth]{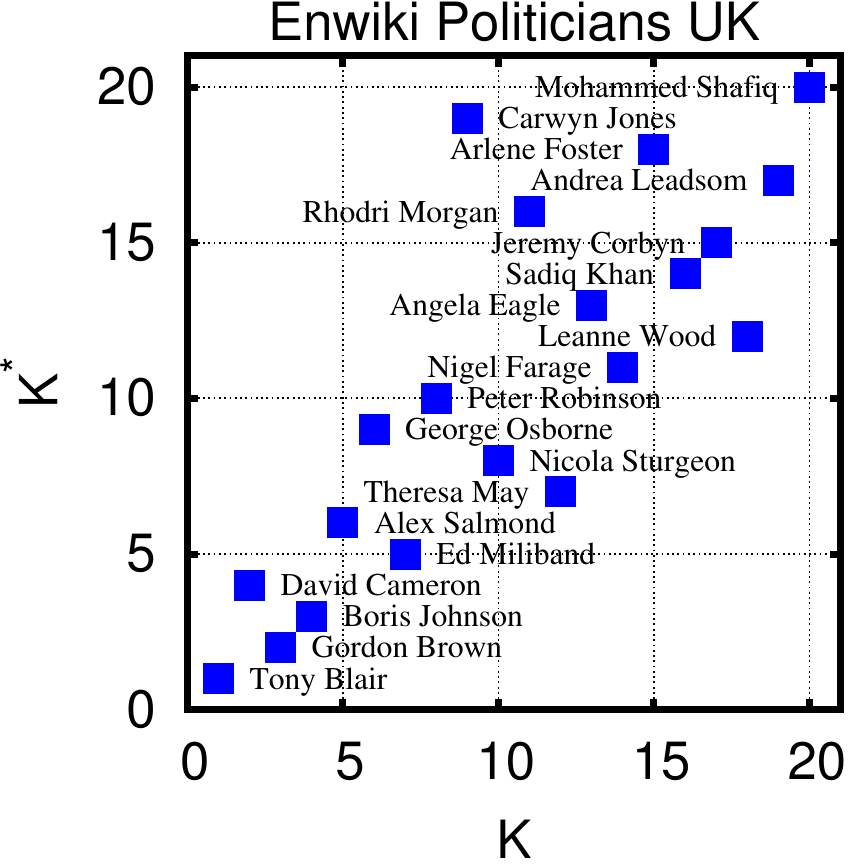}
\caption{Position of nodes in the local $(K,K^*)$ plane of the 
reduced network for 20 UK politicians in the Enwiki network. 
The names are shown on the same lines of the corresponding data points.
}
\label{fig3}
\end{center}
\end{figure}
%%%%%%%%%%%%%%%%%%%%%%%%%%%%%%%%%%%%%%%%%%%%%%%%%%%%%

%%%%%%%%%%%%%%%%%%%%%%%%%%%%%%%%%%%%%%%%%%%%%%%%%%%%%
\begin{table}
\begin{center}
\begin{tabular}{|l|r|r|r|}
\hline
Names (UK) & $K$ & $K^*$ & $K_G$ \\
\hline
\hline
Tony Blair & 1 & 1 & 2 \\
\hline
David Cameron & 2 & 4 & 3 \\
\hline
Gordon Brown & 3 & 2 & 1 \\
\hline
Boris Johnson & 4 & 3 & 7 \\
\hline
Alex Salmond & 5 & 6 & 5 \\
\hline
George Osborne & 6 & 9 & 6 \\
\hline
Ed Miliband & 7 & 5 & 4 \\
\hline
Peter Robinson  & 8 & 10 & 12 \\
\hline
Carwyn Jones & 9 & 19 & 13 \\
\hline
Nicola Sturgeon & 10 & 8 & 9 \\
\hline
Rhodri Morgan & 11 & 16 & 16 \\
\hline
Theresa May & 12 & 7 & 8 \\
\hline
Angela Eagle & 13 & 13 & 15 \\
\hline
Nigel Farage & 14 & 11 & 10 \\
\hline
Arlene Foster & 15 & 18 & 14 \\
\hline
Sadiq Khan & 16 & 14 & 11 \\
\hline
Jeremy Corbyn & 17 & 15 & 17 \\
\hline
Leanne Wood & 18 & 12 & 18 \\
\hline
Andrea Leadsom & 19 & 17 & 19 \\
\hline
Mohammed Shafiq & 20 & 20 & 20 \\
\hline
\end{tabular}

\caption{Same as Table \ref{tab2} for 20 selected UK politicians 
and the English Wikipedia edition of 2013.}
\label{tab3}
\end{center}
\end{table}
%%%%%%%%%%%%%%%%%%%%%%%%%%%%%%%%%%%%%%%%%%%%%%%%%%%%%

\begin{figure}
\begin{center}
\includegraphics[width=0.48\textwidth]{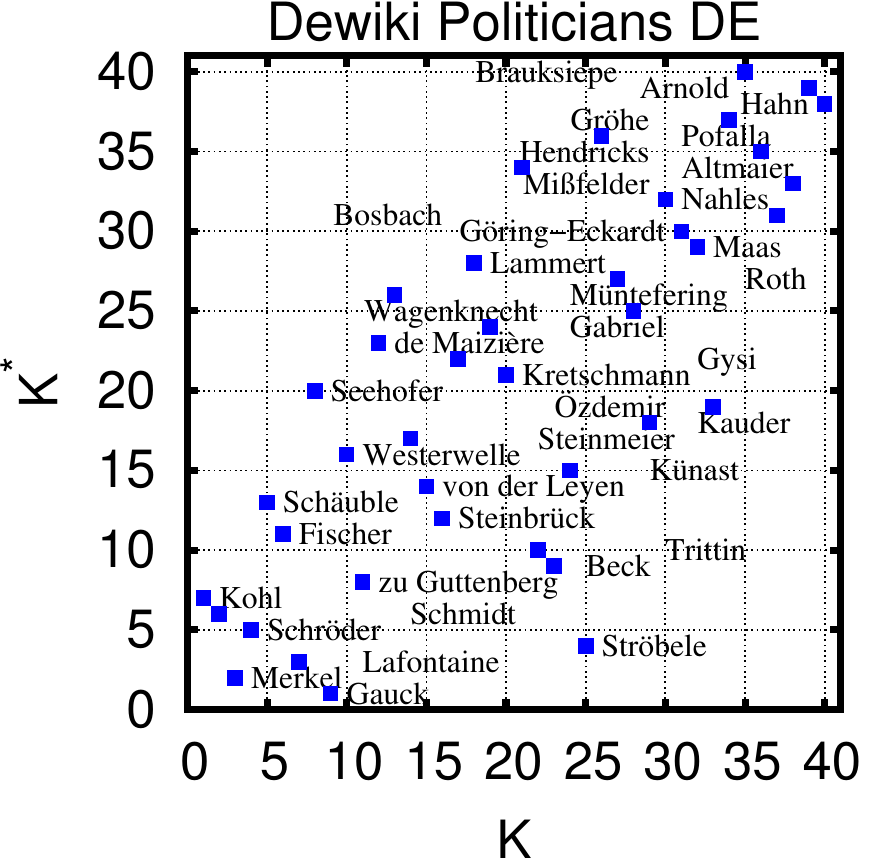}
\caption{Position of nodes in the local $(K,K^*)$ plane of the 
reduced network for 40 DE politicians in the Dewiki network.
The names are shown on the same lines of the corresponding data points. 
Some of the names are horizontally shifted for a better visibility. 
}
\label{fig4}
\end{center}
\end{figure}

\begin{figure}
\begin{center}
\includegraphics[width=0.48\textwidth]{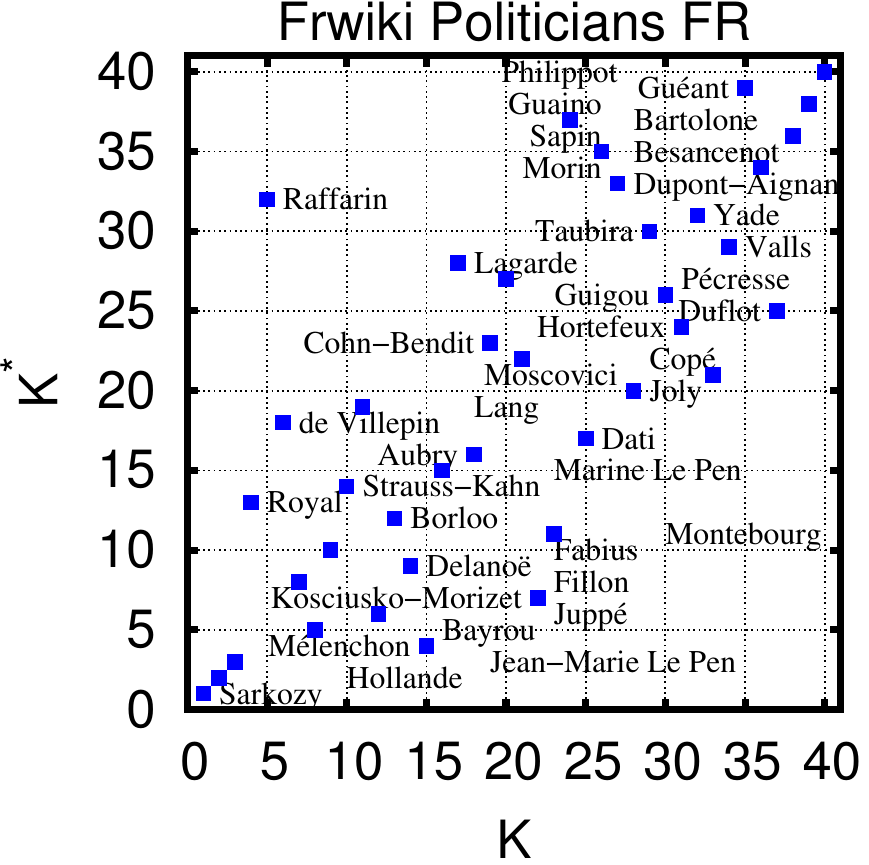}
\caption{Position of nodes in the local $(K,K^*)$ plane of the 
reduced network for 40 FR politicians in the Frwiki network.
The names are shown on the same lines of the corresponding data points.
Some of the names are horizontally shifted for a better visibility. 
}
\label{fig5}
\end{center}
\end{figure}

\begin{figure}
\begin{center}
\includegraphics[width=0.48\textwidth]{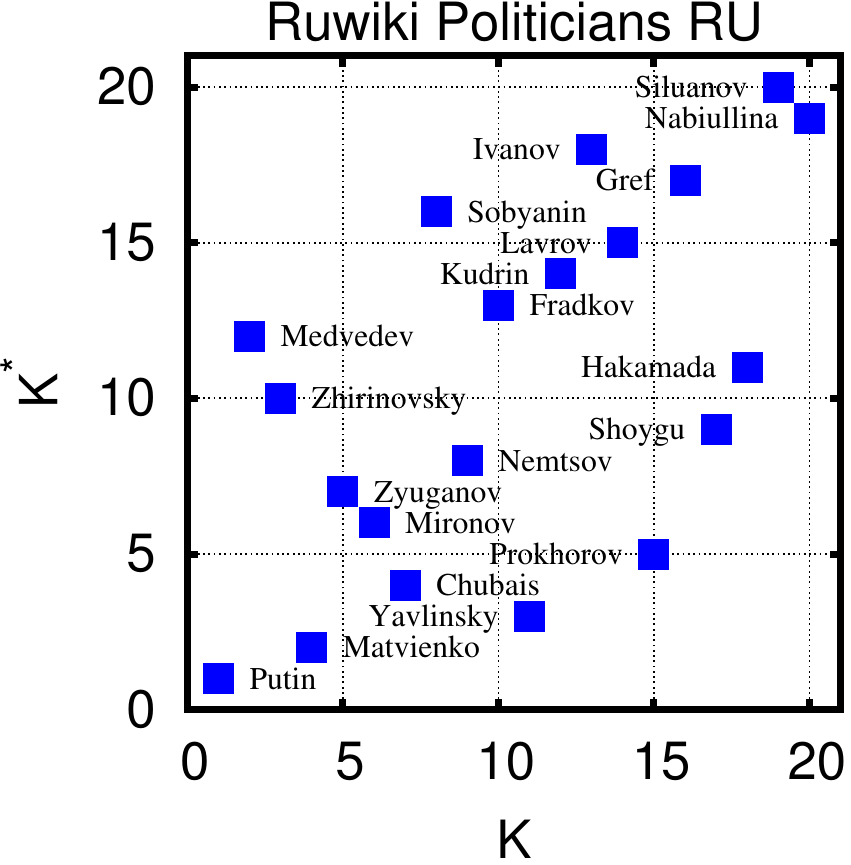}
\caption{Position of nodes in the local $(K,K^*)$ plane of the 
reduced network for 20 RU politicians in the Ruwiki network.
The names are shown on the same lines of the corresponding data points.
}
\label{fig6}
\end{center}
\end{figure}

\begin{figure}
\begin{center}
\includegraphics[width=0.48\textwidth]{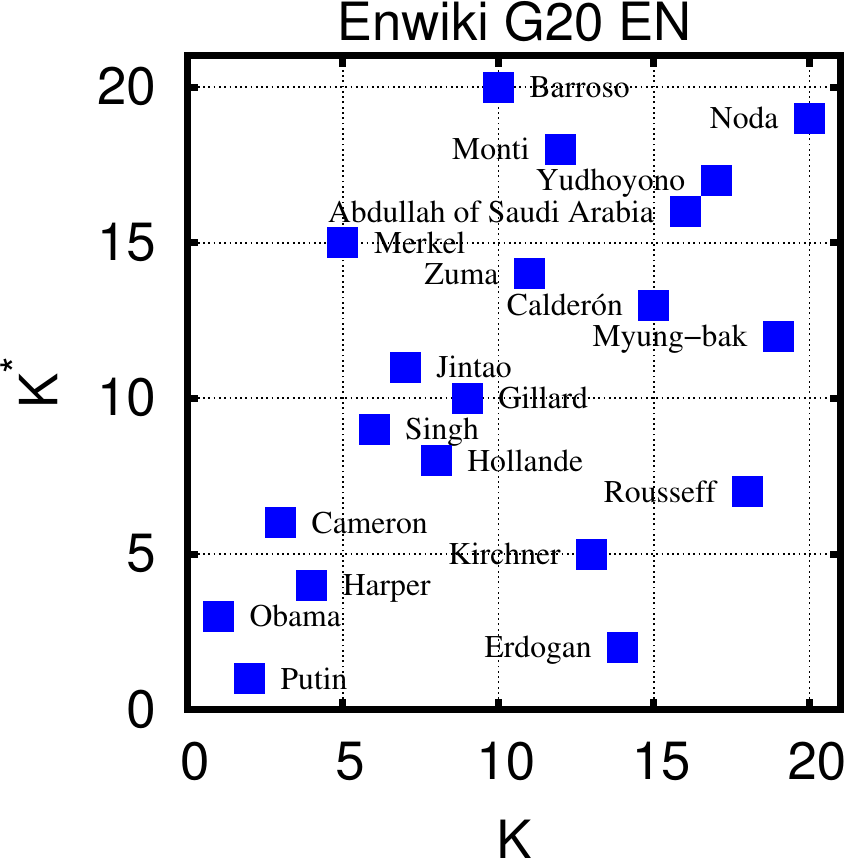}
\caption{Position of nodes in the local $(K,K^*)$ plane of the 
reduced network for 20 state leaders of G20 states in the Enwiki network.
The names are shown on the same lines of the corresponding data points.
}
\label{fig7}
\end{center}
\end{figure}

For each group (or subset of $N_r$ nodes) 
we order politicians by their PageRank probability
in the corresponding global Wikipedia network. After such ordering
we obtain local rank PageRank index $K$
changing from $1$ to $20$ (or $40$). The best known politicians
are found to be at the top values $K=1,2,...$.
In addition we determine the local CheiRank index $K^*$
of the selected names using the CheiRank vector
of the global network. At the top of $K^*$ we have
most communicative articles of politicians.
Then we present the distribution of politicians or G20 state leaders on 
the PageRank-CheiRank plane of local indexes $(K,K^*)$
in Figs.~\ref{fig2}, \ref{fig3}, \ref{fig4}, \ref{fig5}, \ref{fig6} 
and \ref{fig7} for
US, UK, DE, FR, RU and G20 respectively. The full names of political 
leaders are given respectively in 
Tables~\ref{tab2}, \ref{tab3}, \ref{tab4}, \ref{tab5}, \ref{tab6} 
and \ref{tab7} 
with corresponding values of local $K,K^*$ indexes 
(we discuss the meaning of the additional index $K_G$ later).

%%%%%%%%%%%%%%%%%%%%%%%%%%%%%%%%%%%%%%%%%%%%%%%%%%%%%
%%%%%%%%%%%%%%%%%%%%%%%%%%%%%%%%%%%%%%%%%%%%%%%%%%%%%

%%%%%%%%%%%%%%%%%%%%%%%%%%%%%%%%%%%%%%%%%%%%%%%%%%%%%
\begin{table}
\begin{center}
\begin{tabular}{|l|r|r|r|}
\hline
Names (DE) & $K$ & $K^*$ & $K_G$ \\
\hline
\hline
Helmut Kohl & 1 & 7 & 3 \\
\hline
Helmut Schmidt & 2 & 6 & 6 \\
\hline
Angela Merkel & 3 & 2 & 1 \\
\hline
Gerhard Schröder & 4 & 5 & 2 \\
\hline
Wolfgang Schäuble & 5 & 13 & 9 \\
\hline
Joschka Fischer & 6 & 11 & 7 \\
\hline
Oskar Lafontaine & 7 & 3 & 4 \\
\hline
Horst Seehofer & 8 & 20 & 20 \\
\hline
Joachim Gauck & 9 & 1 & 24 \\
\hline
Guido Westerwelle & 10 & 16 & 14 \\
\hline
Karl-Theodor zu Guttenberg & 11 & 8 & 13 \\
\hline
Thomas de Maizière & 12 & 23 & 16 \\
\hline
Franz Müntefering & 13 & 26 & 5 \\
\hline
Frank-Walter Steinmeier & 14 & 17 & 10 \\
\hline
Ursula von der Leyen & 15 & 14 & 25 \\
\hline
Peer Steinbrück & 16 & 12 & 8 \\
\hline
Gregor Gysi & 17 & 22 & 17 \\
\hline
Norbert Lammert & 18 & 28 & 22 \\
\hline
Sigmar Gabriel & 19 & 24 & 12 \\
\hline
Winfried Kretschmann & 20 & 21 & 35 \\
\hline
Peter Altmaier & 21 & 34 & 29 \\
\hline
Jürgen Trittin & 22 & 10 & 15 \\
\hline
Volker Beck & 23 & 9 & 32 \\
\hline
Renate Künast & 24 & 15 & 23 \\
\hline
Hans-Christian Ströbele & 25 & 4 & 11 \\
\hline
Ronald Pofalla & 26 & 36 & 19 \\
\hline
Claudia Roth & 27 & 27 & 21 \\
\hline
Sahra Wagenknecht & 28 & 25 & 28 \\
\hline
Volker Kauder & 29 & 18 & 34 \\
\hline
Andrea Nahles & 30 & 32 & 18 \\
\hline
Katrin Göring-Eckardt & 31 & 30 & 26 \\
\hline
Heiko Maas & 32 & 29 & 30 \\
\hline
Cem Özdemir & 33 & 19 & 33 \\
\hline
Hermann Gröhe & 34 & 37 & 31 \\
\hline
Ralf Brauksiepe & 35 & 40 & 38 \\
\hline
Barbara Hendricks  & 36 & 35 & 36 \\
\hline
Wolfgang Bosbach & 37 & 31 & 27 \\
\hline
Philipp Mißfelder & 38 & 33 & 37 \\
\hline
Rainer Arnold & 39 & 39 & 39 \\
\hline
Florian Hahn & 40 & 38 & 40 \\
\hline
\end{tabular}

\caption{Same as Table \ref{tab2} for 40 selected DE politicians 
and the German Wikipedia edition of 2013.}
\label{tab4}
\end{center}
\end{table}
%%%%%%%%%%%%%%%%%%%%%%%%%%%%%%%%%%%%%%%%%%%%%%%%%%%%%

For the US case shown in Fig.~\ref{fig2} and Table~\ref{tab2}
we find that Obama, Bush and Clinton take the top
three $K$ positions which appears to be rather natural. 
However, the most communicative politicians being at the
top of CheiRank with $K^*=1, 2, 3$
are McCain, Bush and Bloom\-berg.

In Fig.~\ref{fig3} and Table~\ref{tab3}
the names and distribution of 20 UK politicians are
shown. The top 3 positions of PageRank
are taken by UK prime ministers Blair, Cameron, Brown
with $K=1,2,3$. The distribution in the $(K,K^*)$
plane is more centered in a diagonal vicinity as 
compared to the US case and other countries discussed below.
For $K^*=1,2,3$ we have Blair, Brown and Johnson.
The present prime minister May is rather far in rank indexes.

The PageRank-CheiRank distribution of German politicians 
is shown in Fig.~\ref{fig4} with the full names
and ranks given in Table~\ref{tab4}. 
The top PageRank values $K=1,2,3$ are taken by
the chancellors
Kohl, Schmidt and Merkel with
Schr\"oder at $K=4$. However, the most communicative
politicians are Gauck (DE president, $K^*=1$),
Merkel ($K^*=2$) and Lafontaine ($K^*=3$).

%%%%%%%%%%%%%%%%%%%%%%%%%%%%%%%%%%%%%%%%%%%%%%%%%%%%%
%%%%%%%%%%%%%%%%%%%%%%%%%%%%%%%%%%%%%%%%%%%%%%%%%%%%%

%%%%%%%%%%%%%%%%%%%%%%%%%%%%%%%%%%%%%%%%%%%%%%%%%%%%%
\begin{table}
\begin{center}
\begin{tabular}{|l|r|r|r|}
\hline
Names (FR) & $K$ & $K^*$ & $K_G$ \\
\hline
\hline
{\color{blue} Nicolas Sarkozy CB} & 1 & 1 & 1 \\
\hline
{\color{magenta} François Hollande CM} & 2 & 2 & 2 \\
\hline
{\color{blue-violet} Jean-Marie Le Pen CV} & 3 & 3 & 11 \\
\hline
{\color{magenta} Ségolène Royal CM} & 4 & 13 & 3 \\
\hline
{\color{blue} Jean-Pierre Raffarin CB} & 5 & 32 & 10 \\
\hline
{\color{blue} Dominique de Villepin CB} & 6 & 18 & 12 \\
\hline
{\color{blue} François Fillon CB} & 7 & 8 & 4 \\
\hline
{\color{blue} François Bayrou CB} & 8 & 5 & 6 \\
\hline
{\color{magenta} Laurent Fabius CM} & 9 & 10 & 8 \\
\hline
{\color{magenta} Dominique Strauss-Kahn CM} & 10 & 14 & 7 \\
\hline
{\color{magenta} Jack Lang CM} & 11 & 19 & 13 \\
\hline
{\color{blue} Alain Juppé CB} & 12 & 6 & 5 \\
\hline
{\color{blue} Jean-Louis Borloo CB} & 13 & 12 & 19 \\
\hline
{\color{magenta} Bertrand Delanoë CM} & 14 & 9 & 18 \\
\hline
{\color{red} Jean-Luc Mélenchon CR} & 15 & 4 & 14 \\
\hline
{\color{blue-violet} Marine Le Pen CV} & 16 & 15 & 17 \\
\hline
{\color{blue} Christine Lagarde CB} & 17 & 28 & 25 \\
\hline
{\color{magenta} Martine Aubry CM} & 18 & 16 & 9 \\
\hline
{\color{ao(english)} Daniel Cohn-Bendit CG} & 19 & 23 & 32 \\
\hline
{\color{blue} Valérie Pécresse CB} & 20 & 27 & 29 \\
\hline
{\color{blue} Jean-François Copé CB} & 21 & 22 & 15 \\
\hline
{\color{blue} Nathalie Kosciusko-Morizet CB} & 22 & 7 & 26 \\
\hline
{\color{magenta} Arnaud Montebourg CM} & 23 & 11 & 16 \\
\hline
{\color{magenta} Claude Bartolone CM} & 24 & 37 & 28 \\
\hline
{\color{blue} Rachida Dati CB} & 25 & 17 & 35 \\
\hline
{\color{red} Olivier Besancenot CR} & 26 & 35 & 37 \\
\hline
{\color{blue-violet} Nicolas Dupont-Aignan CV} & 27 & 33 & 30 \\
\hline
{\color{ao(english)} Eva Joly CG} & 28 & 20 & 21 \\
\hline
{\color{magenta} Christiane Taubira CM} & 29 & 30 & 38 \\
\hline
{\color{magenta} Élisabeth Guigou CM} & 30 & 26 & 23 \\
\hline
{\color{blue} Brice Hortefeux CB} & 31 & 24 & 27 \\
\hline
{\color{blue} Rama Yade CB} & 32 & 31 & 39 \\
\hline
{\color{magenta} Pierre Moscovici CM} & 33 & 21 & 20 \\
\hline
{\color{magenta} Manuel Valls CM} & 34 & 29 & 22 \\
\hline
{\color{blue} Claude Guéant CB} & 35 & 39 & 34 \\
\hline
{\color{blue} Hervé Morin CB} & 36 & 34 & 24 \\
\hline
{\color{ao(english)} Cécile Duflot CG} & 37 & 25 & 36 \\
\hline
{\color{magenta} Michel Sapin CM} & 38 & 36 & 31 \\
\hline
{\color{blue} Henri Guaino CB} & 39 & 38 & 33 \\
\hline
{\color{blue-violet} Florian Philippot CV} & 40 & 40 & 40 \\
\hline
\end{tabular}

\caption{Same as Table \ref{tab2} for 40 selected FR politicians 
and the French Wikipedia edition of 2013. Color marks membership 
in political parties: red CR (far-left parties),
magenta CM (socialist party PS), green CG (green parties),
blue CB (right parties UMP, UDI), CV violet (far-right and FN).}
\label{tab5}
\end{center}
\end{table}
%%%%%%%%%%%%%%%%%%%%%%%%%%%%%%%%%%%%%%%%%%%%%%%%%%%%%

The French politicians are presented in Fig.~\ref{fig5}
and Table~\ref{tab5}. Here we choose names of those
who are really active in the period 2007 - 2013, 
thus e.~g. Jacques Chirac is not included in the list.
The top 3 positions in $K$ and $K^*$ are taken
by two presidents Sarkozy, Hollande with
(Jean-Marie) Le Pen at the third position. We note a large dispersion
of positions on the $(K,K^*)$ plane for the main
part of politicians. Thus Royal ($K=4$) 
and Raffarin ($K=5$) have rather high values of $K^* =13$ or 32
respectively. For French politicians in Table~\ref{tab5}
we mark by color their membership in political
parties. The effect of interactions and links between different
parties is discussed in the last Section.

%%%%%%%%%%%%%%%%%%%%%%%%%%%%%%%%%%%%%%%%%%%%%%%%%%%%%
%%%%%%%%%%%%%%%%%%%%%%%%%%%%%%%%%%%%%%%%%%%%%%%%%%%%%

%%%%%%%%%%%%%%%%%%%%%%%%%%%%%%%%%%%%%%%%%%%%%%%%%%%%%
\begin{table}
\begin{center}
\begin{tabular}{|l|r|r|r|}
\hline
Names (RU) & $K$ & $K^*$ & $K_G$ \\
\hline
\hline
Putin, Vladimir Vladimirovich & 1 & 1 & 1 \\
\hline
Medvedev, Dmitry Anatolyevich & 2 & 12 & 2 \\
\hline
Zhirinovsky, Vladimir Wolfovich & 3 & 10 & 15 \\
\hline
Matvienko, Valentina Ivanovna & 4 & 2 & 11 \\
\hline
Zyuganov, Gennady Andreyevich & 5 & 7 & 13 \\
\hline
Mironov, Sergey Mikhailovich & 6 & 6 & 8 \\
\hline
Chubais, Anatoly Borisovich & 7 & 4 & 6 \\
\hline
Sobyanin, Sergey Semenovich & 8 & 16 & 9 \\
\hline
Nemtsov, Boris Yefimovich & 9 & 8 & 10 \\
\hline
Fradkov, Mikhail Yefimovich & 10 & 13 & 3 \\
\hline
Yavlinsky, Grigory & 11 & 3 & 14 \\
\hline
Kudrin, Alexei Leonidovich & 12 & 14 & 7 \\
\hline
Ivanov, Sergey Borisovich & 13 & 18 & 4 \\
\hline
Lavrov, Sergey Viktorovich & 14 & 15 & 20 \\
\hline
Prokhorov, Mikhail Dmitrievich & 15 & 5 & 18 \\
\hline
Gref, Herman Oskarovich & 16 & 17 & 5 \\
\hline
Shoygu, Sergei & 17 & 9 & 19 \\
\hline
Hakamada, Irina Mutsuovna & 18 & 11 & 17 \\
\hline
Siluanov, Anton Germanovich & 19 & 20 & 16 \\
\hline
Nabiullina, Elvira & 20 & 19 & 12 \\
\hline
\end{tabular}

\caption{Same as Table \ref{tab2} for 20 selected RU politicians 
and the Russian Wikipedia edition of 2013.}
\label{tab6}
\end{center}
\end{table}
%%%%%%%%%%%%%%%%%%%%%%%%%%%%%%%%%%%%%%%%%%%%%%%%%%%%%

The names of 20 Russian politicians and their
distribution on the $(K,K^*)$ plane are presented in Fig.~\ref{fig6}
and Table~\ref{tab6}. Similar to the case of France
the two top PageRank positions are taken by presidents
Putin, Medvedev with Zhirinovsky at the third position with $K=3$.
For $K^*=1,2,3$ we have Putin, Matvienko, Yavlinsky
while Medvedev is at the far value $K^*=12$ showing
low communicative properties of his Wikipedia article.

%%%%%%%%%%%%%%%%%%%%%%%%%%%%%%%%%%%%%%%%%%%%%%%%%%%%%
%%%%%%%%%%%%%%%%%%%%%%%%%%%%%%%%%%%%%%%%%%%%%%%%%%%%%

%%%%%%%%%%%%%%%%%%%%%%%%%%%%%%%%%%%%%%%%%%%%%%%%%%%%%
\begin{table}
\begin{center}
\begin{tabular}{|l|r|r|r|r|}
\hline
G20 / Wikipedia edition & EN & DE & FR & RU \\
\hline
\hline
Names & $K$ & $K$ & $K$ & $K$ \\
\hline
\hline
Barack Obama & 1 & 1 & 1 & 2 \\
\hline
Vladimir Putin & 2 & 3 & 2 & 1 \\
\hline
David Cameron & 3 & 4 & 6 & 4 \\
\hline
Stephen Harper & 4 & 10 & 7 & 16 \\
\hline
Angela Merkel & 5 & 2 & 4 & 3 \\
\hline
Manmohan Singh & 6 & 13 & 13 & 10 \\
\hline
Hu Jintao & 7 & 9 & 5 & 5 \\
\hline
François Hollande & 8 & 5 & 3 & 6 \\
\hline
Julia Gillard & 9 & 14 & 16 & 11 \\
\hline
José Manuel Barroso & 10 & 6 & 8 & 14 \\
\hline
Jacob Zuma & 11 & 18 & 18 & 18 \\
\hline
Mario Monti & 12 & 8 & 9 & 9 \\
\hline
Cristina Fernández de Kirchner & 13 & 12 & 10 & 8 \\
\hline
Recep Tayyip Erdoğan & 14 & 7 & 14 & 12 \\
\hline
Felipe Calderón & 15 & 11 & 12 & 7 \\
\hline
Abdullah of Saudi Arabia & 16 & 15 & 15 & 13 \\
\hline
Susilo Bambang Yudhoyono & 17 & 19 & 20 & 19 \\
\hline
Dilma Rousseff & 18 & 17 & 11 & 15 \\
\hline
Lee Myung-bak & 19 & 16 & 17 & 17 \\
\hline
Yoshihiko Noda & 20 & 20 & 19 & 20 \\
\hline
\end{tabular}

\caption{List of the 20 state leaders of the G20 states and the 
corresponding PageRank index values $K$ of the reduced network 
matrices $G_{\rm R}$ for the English (EN), German (DE), French (FR) 
and Russian (RU) Wikipedia editions of 2013.
For each state leader the country name is given in ISO 3166-1 alpha-2.}
\label{tab7}
\end{center}
\end{table}
%%%%%%%%%%%%%%%%%%%%%%%%%%%%%%%%%%%%%%%%%%%%%%%%%%%%%

All the groups of politicians have been
considered above in the frame of Wikipedia editions
corresponding to their native language. Indeed, 
we find that in other language editions 
the articles about many politicians are rather short
or sometimes they are even absent
(e.~g. for 40 French politicians
in Enwiki or Ruwiki).
However, the political leaders of countries are
usually well present in the editions
discussed here. Therefore,
we take for our analysis 20 world political leaders
that have participated in the G20 meeting at Los Cabos summit
in 2012 \cite{g20wiki}. Their names and 
local PageRank indexes
according to Enwiki, Dewiki, Frwiki, Ruwiki
are given in Table~\ref{tab7}. The distribution
of politicians on PageRank-CheiRank plane
is shown in Fig.~\ref{fig7}
for Enwiki. We take the name of country leader
Abdullah of Saudi Arabia (Saudi Arabia)
since the name of the minister of finance,
who was representing Saudi Arabia,
is not listed in Dewiki, Frwiki, Ruwiki;
Jos\'e Manuel Barroso is taken as EU representative.

Among G20 leaders the top 2 PageRank positions are taken
by Obama and Putin (see Table~\ref{tab7}) in Enwiki, Frwiki;
Putin and Obama in Ruwiki and Obama and Merkel in Dewiki.
So there is a definite trend for leaders being promoted in their
native editions. The language preference is
probably the reason to have Singh (India)
ahead of Jintao (China) in Enwiki
while in other editions Jintao is well ahead of Singh.
At the top CheiRank positions $K^*$ of Enwiki we have
Putin, Erdoğan, Obama (see Fig.~\ref{fig7})
showing very different communicative strengths of 
political leaders. 

In the next sections we consider 
interactions between selected political leaders
using the reduced Google matrix approach.

\section{Direct and hidden links of US politicians}

\begin{figure}
\begin{center}
\includegraphics[width=0.48\textwidth]{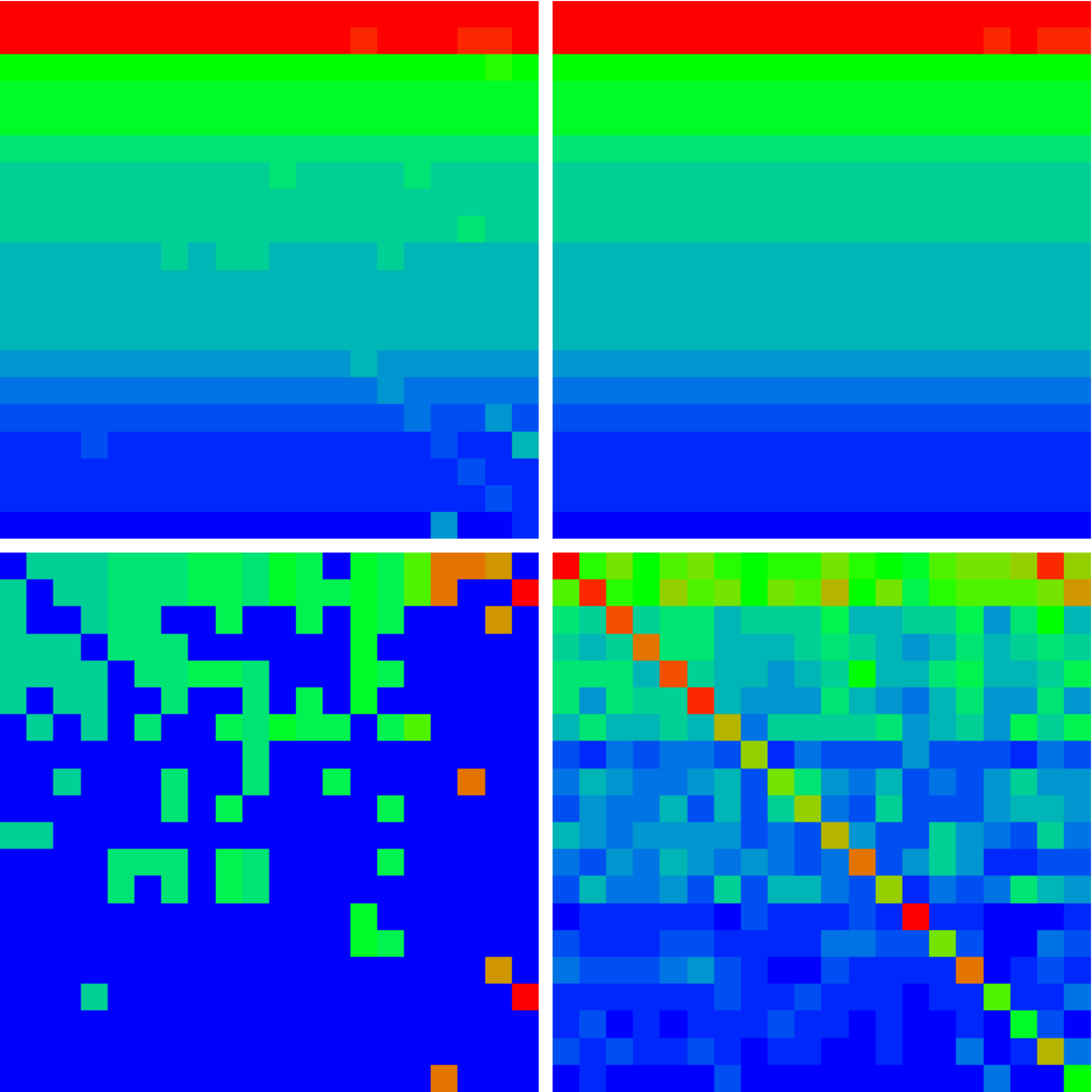}
\caption{Density plots of the matrices $G_{\rm R}$ (top left), 
$G_{\rm pr}$ (top right), $G_{\rm rr}$ (bottom left) and $G_{\rm qr}$ 
(bottom right) for the reduced network of 20 US politicians 
in the Enwiki network. The nodes $N_r$ are ordered in lines by
PageRank index changing from $K=1$ to $N_r$ (left to right)
and in columns from $K'=1$ to $N_r$ from top to bottom.
The weights of the three matrix components of $G_{\rm R}$ are  
$W_{\rm pr}=0.9640$ for $G_{\rm pr}$, $W_{rr} = 0.01970$ for $G_{rr}$,
$W_{\rm qr}= 0.01630$ for $G_{qr}$. 
The colors represent maximum (red), intermediate 
(green) and minimum (blue) values. 
}
\label{fig8}
\end{center}
\end{figure}

The reduced Google matrix $G_{\rm R}$ of 20 US politicians,
listed in Fig.~\ref{fig2}  and Table~\ref{tab2},
is shown in Fig.~\ref{fig8} with its three
matrix components $G_{\rm pr}$, $G_{rr}$, $G_{\rm qr}$
from (\ref{eq_3terms}). The amplitudes of matrix elements
are shown by color with maximum for red
and minimum for blue. We mention that for the data 
of US Polititicians in this section there are no negative matrix elements of 
$G_{\rm qr}$. The same holds for the cases of DE and FR politicians 
(see sections below). However for UK politicians (G20 data, also see sections 
below for both cases) there are few very (rather) small 
negative elements of $G_{\rm qr}$ of order $\sim 10^{-8}$ (or $\sim 10^{-5}$) 
with no (or very small) effects on colors. The data 
of US politicians in Fig.~\ref{fig8} clearly show that 
the main contribution to $G_{\rm R}$ is given by
the projector component $G_{\rm pr}$ with a weight of 
approximately $96$\%. The remaining weight is
distributed between the component of direct links
$G_{rr}$ ($1.9$\%) and the one 
of indirect links $G_{\rm qr}$ ($1.6$\%) (see Fig.~\ref{fig8}).
Of course, the total weight of 
three components is equal to unity by construction.

%%%%%%%%%%%%%%%%%%%%%%%%%%%%%%%%%%%%%%%%%%%%%%%%%%%%%
%%%%%%%%%%%%%%%%%%%%%%%%%%%%%%%%%%%%%%%%%%%%%%%%%%%%%

Thus the main component $G_{\rm pr}$ imposes to $G_{\rm R}$ 
a large contribution proportional to 
the PageRank probability $P(K)$ which is mainly produced by
the environment of the huge remaining part $G_{ss}$ of 
the global network with $N-N_r \gg N_r$ nodes.
Due to this the structure of $G_{\rm R}$ is 
close to a solution (discussed above)
when each column of $G_{\rm R}$ is roughly 
given by the same PageRank vector of $N_r$
nodes which up to a constant factor coincides 
with the PageRank probabilities of the 
selected $N_r$ nodes in the global network of $N$ nodes. 
Due to the simple rank-one structure of $G_{\rm pr}$ 
the smaller contributions of the other two components
$G_{rr}$ of direct 
and $G_{\rm qr}$ of indirect links (friends)
play an important role even if their weight is
significantly smaller as compared to
the weight of $G_{\rm pr}$.

%%%%%%%%%%%%%%%%%%%%%%%%%%%%%%%%%%%%%%%%%%%%%%%%%%%%%
\begin{figure}
\begin{center}
\includegraphics[width=0.48\textwidth]{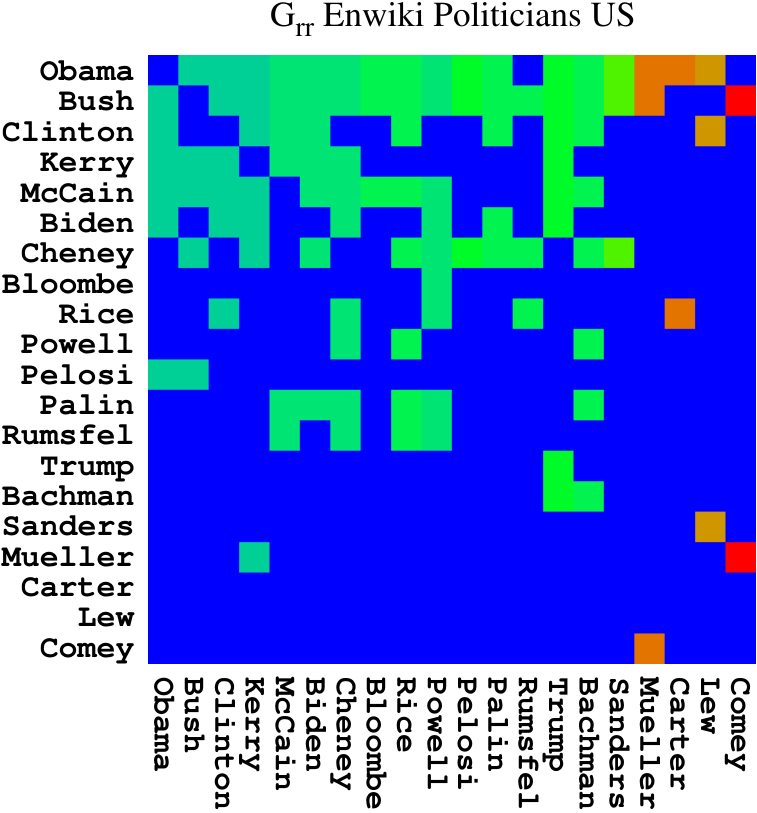}
\caption{Density plot of the matrix $G_{rr}$ for the reduced network 
of 20 US politicians in the Enwiki network with short names at both axes
(up to 7 letters, full names are in Table~\ref{tab2}).}
\label{fig9}
\end{center}
\end{figure}
%%%%%%%%%%%%%%%%%%%%%%%%%%%%%%%%%%%%%%%%%%%%%%%%%%%%%

%%%%%%%%%%%%%%%%%%%%%%%%%%%%%%%%%%%%%%%%%%%%%%%%%%%%%
\begin{figure}
\begin{center}
\includegraphics[width=0.48\textwidth]{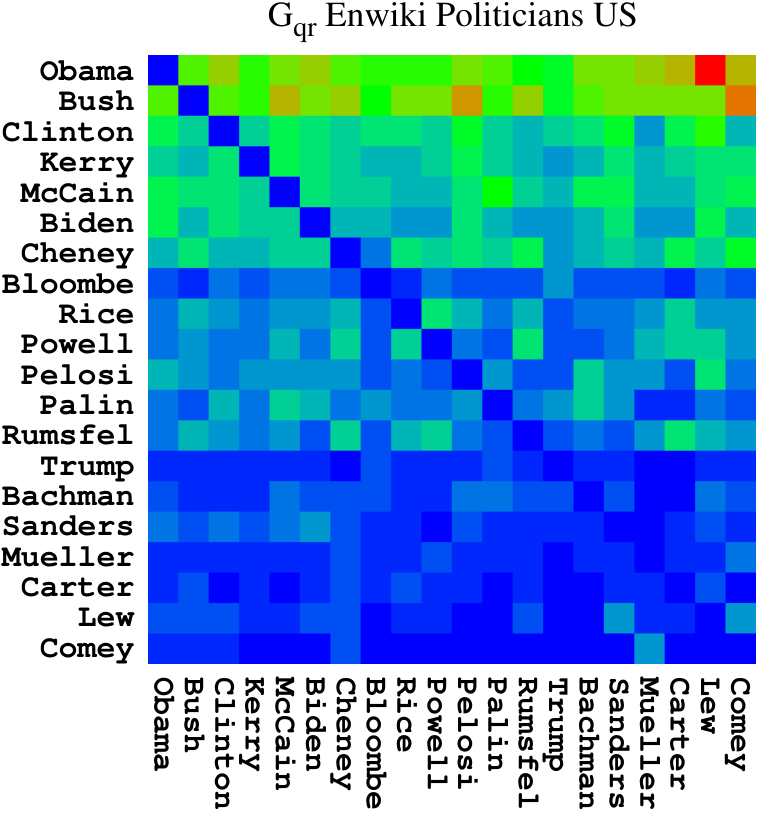}
\caption{Density plot of the matrix $G_{qr}$ without diagonal elements 
for the reduced network of 20 US politicians in the Enwiki network 
with short names at both axes. The weight of this matrix component
without diagonal is $W_{\rm qrnd}= 0.01143$
(see Fig.~\ref{fig8} for $W_{\rm qr}$ weight with diagonal).
}
\label{fig10}
\end{center}
\end{figure}
%%%%%%%%%%%%%%%%%%%%%%%%%%%%%%%%%%%%%%%%%%%%%%%%%%%%%

%%%%%%%%%%%%%%%%%%%%%%%%%%%%%%%%%%%%%%%%%%%%%%%%%%%%%
\begin{table}
\begin{center}
\begin{tabular}{|l|l|l|}
\hline
Politicians & US & Enwiki \\
\hline
\hline
Name & Friends & Followers \\
\hline
\hline
Obama & Bush & Lew \\
 & Clinton & Comey \\
 & Biden & Carter \\
\hline
Bush & Obama & Comey \\
 & Cheney & Pelosi \\
 & McCain & McCain \\
\hline
Clinton & Obama & Lew \\
 & Bush & Sanders \\
 & McCain & Pelosi \\
\hline
Kerry & Obama & Pelosi \\
 & Bush & McCain \\
 & Clinton & Biden \\
\hline
McCain & Bush & Palin \\
 & Obama & Comey \\
 & Clinton & Sanders \\
\hline
\end{tabular}

\caption{List of leading politicians (in PageRank order) of the group of 
US politicians together with the leading ``Friends'' (``Followers'') in 
this group defined by maximal values of the elements of $G_{qr}$ 
corresponding to links ``to'' (``from'') the Friends (Followers) 
for the English Wikipedia edition. Links to/from the same politician, 
corresponding to diagonal elements of $G_{qr}$, are omitted.}
\label{tab8}
\end{center}
\end{table}
%%%%%%%%%%%%%%%%%%%%%%%%%%%%%%%%%%%%%%%%%%%%%%%%%%%%%

The global structure of $G_{rr}$ of 20 US politicians is shown in
Fig.~\ref{fig8} (bottom left panel)
and in more detail in Fig.~\ref{fig9} 
where the lines and columns are marked by short names of politicians
(up to 7 letters). We can say that large matrix elements in 
a column of a given politician can be considered as direct friends
to whom he/she points in his/her Wikipedia article.
However, we should note that by construction of $G$
all elements inside a given column have the same amplitudes
given by a fraction of total number of outgoing links of a given
article which point inside $N_r$ nodes of selected subset.
Thus on the basis of the $G_{rr}$
component it is not possible to say that some 
direct friend (link) of a given politician
is more preferable then another one:
all of them have the same amplitude.
Of course, when we are speaking about a friend 
we simply mean that one politician points to another
but at present we cannot say if this link has a positive or negative 
content. Such a classification would required further extension of our
$G_{\rm R}$ analysis. However, since the PageRank probability is 
on average proportional to the number of ingoing links we can assume that  
the content is mainly positive.

The large amplitudes inside a given line, attributed to a given
politician, mark the followers of this politician
(similar to a situation in Twitter). Thus from Fig.~\ref{fig9}
we can say that the first (strongest) direct follower of Bush is Comey
who is the present FBI director and is closely linked with Bush.
At the same time he has no direct link with Obama.
Of course the direct links are important but only on their own basis
it is not possible to obtain a correct relationship between
selected persons. Indeed, an attempt 
\cite{aragon2012} to make ranking of 
historical figures of Wikipedia considering only links between
their biographical articles gave a rather strange unrealistic result
clearly showing that this approach is not working
(see also discussion in \cite{eomwiki24}).

Due to the above reasons the most interesting matrix component
$G_{\rm qp}$ is the one of indirect links shown in Fig.~\ref{fig8}
(bottom right panel) and in Fig.~\ref{fig10} where diagonal elements have been removed.
The weight $W_{\rm qrd} = 0.00487$ of diagonal components of $G_{\rm qr}$
is approximately twice as small than the weight of the nondiagonal part
with  $W_{\rm qrnd} = 0.01143$ : nondiagonal elements
play a more significant role. One can understand that indirect 
links produce considerable diagonal contributions. However, for 
the analysis of indirect links between {\em different} nodes there are not 
of interest. 

As discussed above the largest matrix elements of $G_{\rm qr}$
in a column of a given politician
give his/her strongest indirect or hidden friends
while those in a line give his/her strongest hidden followers.
The names of top friends and top followers of
top 5 PageRank US politicians are given in Table~\ref{tab8}.
Surprisingly we find that the strongest hidden followers of Obama
are not his colleagues from the democratic party but Lew 
(actual secretary of treasury),
Comey (actual FBI director), 
Carter (actual secretary of defense).
Interestingly, Clinton has as main followers Lew, Sanders, Pelosi
so that already in 2013 
there existed the strong hidden links between Sanders and
Clinton. At the same time the top 3 hidden friends (followers)
of Trump are Obama, Bush, Clinton (Palin, Bloomberg, Clinton)
highlighting hidden links between these two political leaders
who are now fighting for the US presidency (see data at \cite{ourwebpage}).

It is possible to try not to take into account the projector
component and construct a modified reduced Google matrix
${\tilde G}_{\rm R} $ obtained  from $G_{rr}+G_{\rm qrnd}$
by renormalization of each column to unity. Then the PageRank vector
of $ {\tilde G}_{\rm R} $ gives the new ranking of the selected group
with index $K_G$ given in Table~\ref{tab2}. We then see that
the top 3 positions are taken by
Bush, Obama, McCain. This gives a rearrangement of the ranking
which stresses in a stronger way previous presidential teams.
We will see that for other countries the effects can work in 
a different direction (e.~g. for DE). 

\section{Direct and hidden links of UK politicians}

\begin{figure}
\begin{center}
\includegraphics[width=0.48\textwidth]{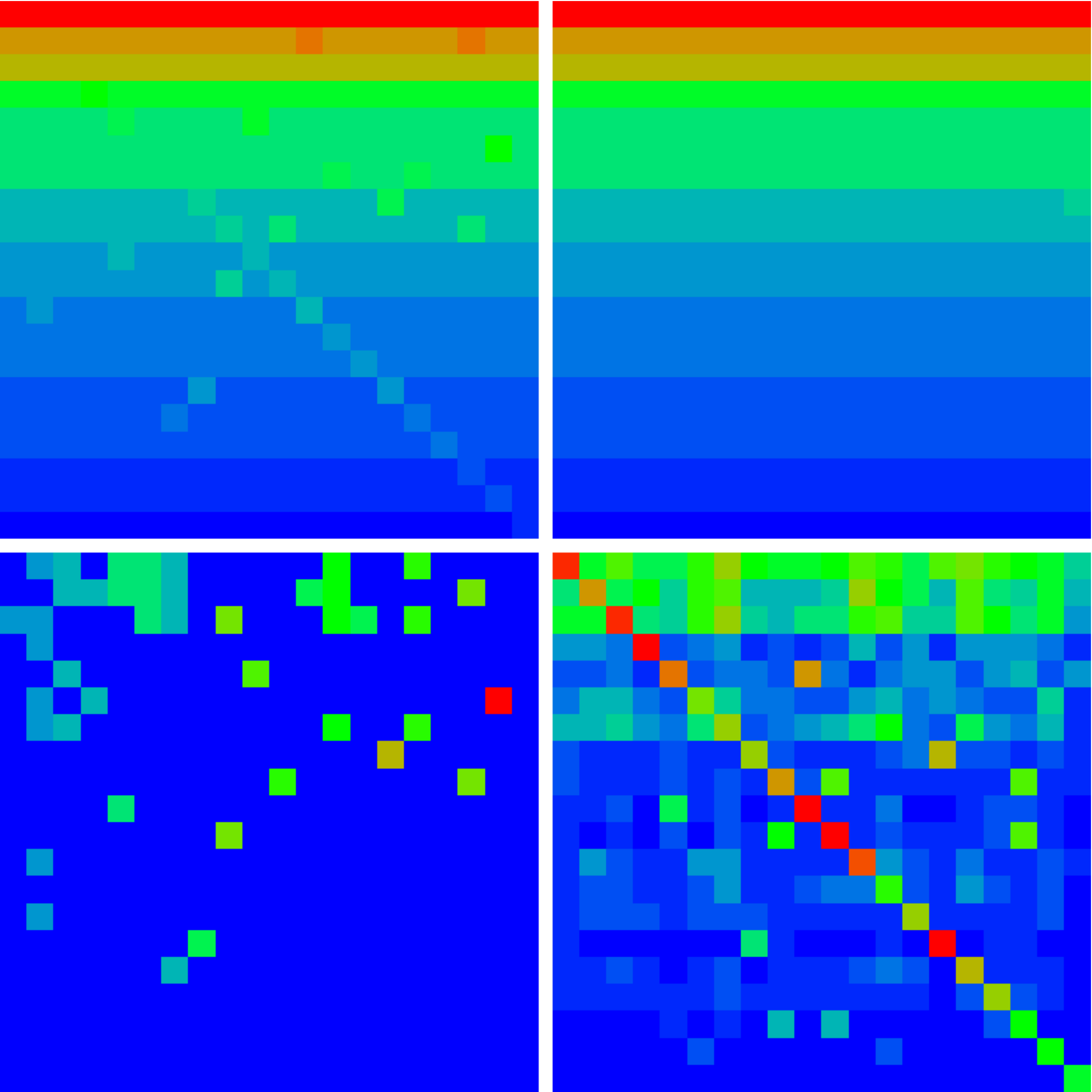}
\caption{Same as Fig. \ref{fig8} for 20 UK politicians in 
the Enwiki network. 
The weights of the three matrix components of $G_{\rm R}$ are  
$W_{\rm pr}=0.9668$ for $G_{\rm pr}$, $W_{rr} = 0.01729$ for $G_{rr}$,
$W_{\rm qr}= 0.01588$ for $G_{qr}$. 
}
\label{fig11}
\end{center}
\end{figure}

\begin{figure}
\begin{center}
\includegraphics[width=0.48\textwidth]{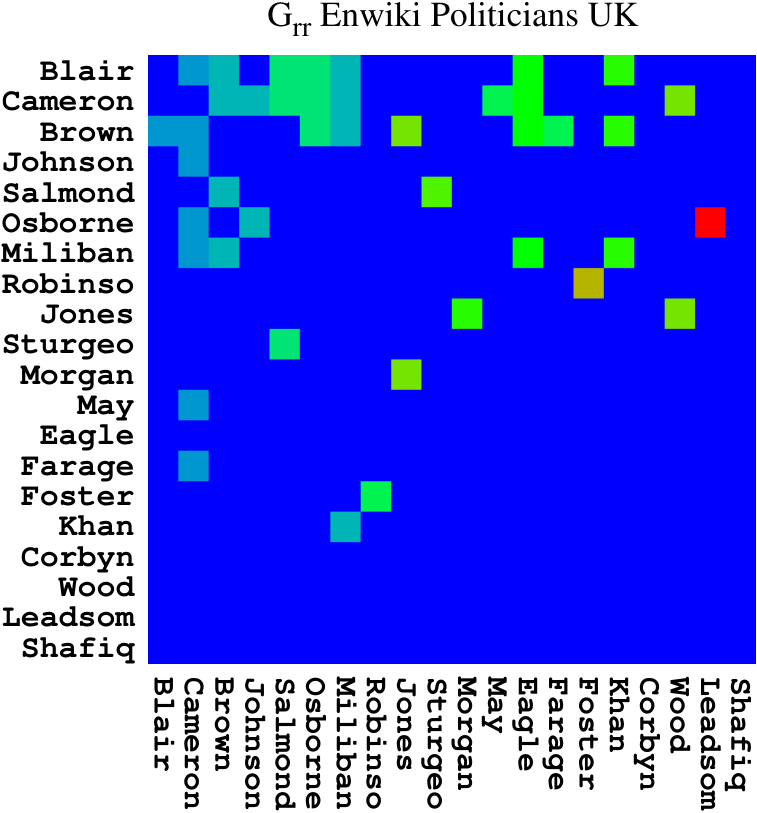}
\caption{Density plot of the matrix $G_{rr}$ for the reduced network 
of 20 UK politicians in the Enwiki network with short names at both axes.}
\label{fig12}
\end{center}
\end{figure}

\begin{figure}
\begin{center}
\includegraphics[width=0.48\textwidth]{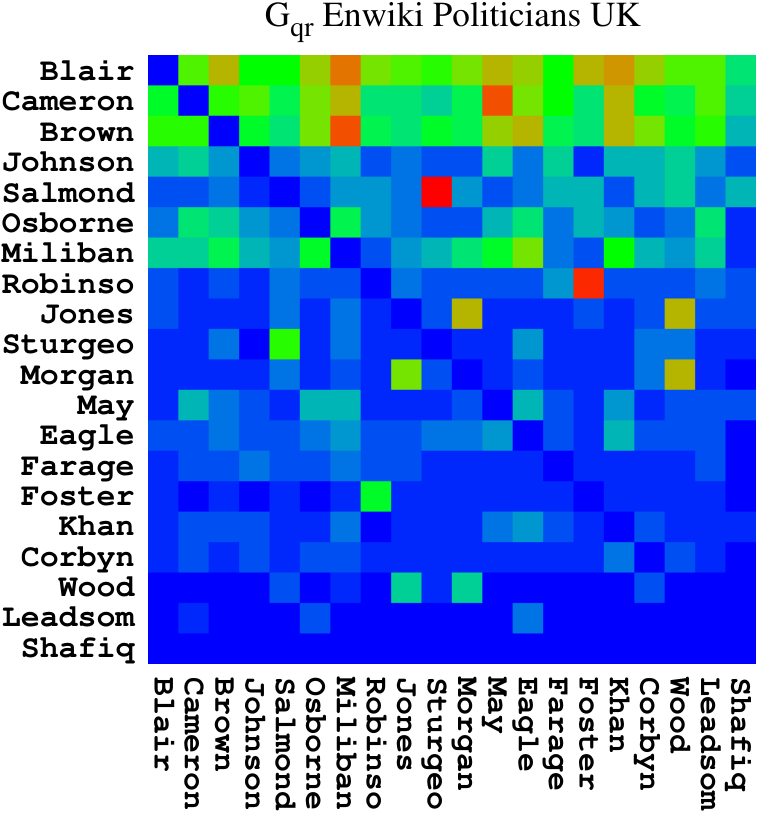}
\caption{Density plot of the matrix $G_{qr}$ without diagonal elements 
for the reduced network of 20 UK politicians in the Enwiki network 
with short names at both axes.
 The weight of this matrix component
without diagonal is $W_{\rm qrnd}= 0.009875$
(see Fig.~\ref{fig11} for $W_{\rm qr}$ weight with diagonal).
}
\label{fig13}
\end{center}
\end{figure}

The reduced Google matrix analysis of 20 UK politicians
from Fig.~\ref{fig3} and Table~\ref{tab3} is presented in
Figs.~\ref{fig11},~\ref{fig12},~\ref{fig13}. As for the US case
we find that $G_{\rm R}$ has the dominant contribution from
the projector component $G_{\rm pr}$ which has a 
similar weight of $96$\%. The weights of other two components
of direct $G_{rr}$ ($1.7$\%) and hidden $G_{\rm qr}$ ($1.5$\%) links
is also similar to those of US. However, on average the direct links
are distributed in a more homogeneous manner for UK then for US
(see Figs.~\ref{fig12},~\ref{fig9}). The strongest direct link is from
Leadsom to Osborne.

%%%%%%%%%%%%%%%%%%%%%%%%%%%%%%%%%%%%%%%%%%%%%%%%%%%%%
%%%%%%%%%%%%%%%%%%%%%%%%%%%%%%%%%%%%%%%%%%%%%%%%%%%%%

The distribution of hidden links in Fig.~\ref{fig13}
is much more broad as compared to the direct ones in Fig.~\ref{fig12}. 
The top 3 hidden friends and followers of top 5 PageRank
politicians are given in Table~\ref{tab9}. There are strong links
between top 3 leaders Blair, Brown and Cameron. More surprisingly
we find that already in 2013, May was the strongest follower of Cameron 
(there is only a moderate direct link between them)
and Johnson (there is no direct link). The strongest amplitudes of links
are from Sturgeon (actual first minister of Scotland)
to Salmond (previous first minister of Scotland) and 
from Foster to Robinson (first ministers of Northern Ireland).
Even if the direct links are present in $G_{rr}$
they are definitely not so pronounced as in the indirect part 
$G_{\rm qr}$. Also we find that Khan (actual mayor of London)
is a second by strength indirect follower of Cameron even if
there are no direct links between them.

%%%%%%%%%%%%%%%%%%%%%%%%%%%%%%%%%%%%%%%%%%%%%%%%%%%%%
%%%%%%%%%%%%%%%%%%%%%%%%%%%%%%%%%%%%%%%%%%%%%%%%%%%%%

%%%%%%%%%%%%%%%%%%%%%%%%%%%%%%%%%%%%%%%%%%%%%%%%%%%%%
%%%%%%%%%%%%%%%%%%%%%%%%%%%%%%%%%%%%%%%%%%%%%%%%%%%%%

%%%%%%%%%%%%%%%%%%%%%%%%%%%%%%%%%%%%%%%%%%%%%%%%%%%%%
\begin{table}
\begin{center}
\begin{tabular}{|l|l|l|}
\hline
Politicians & UK & Enwiki \\
\hline
\hline
Name & Friends & Followers \\
\hline
\hline
Blair & Brown & Miliband \\
 & Cameron & Khan \\
 & Miliband & Foster \\
\hline
Cameron & Blair & May \\
 & Brown & Khan \\
 & Osborne & Miliband \\
\hline
Brown & Blair & Miliband \\
 & Cameron & Khan \\
 & Miliband & Eagle \\
\hline
Johnson & Cameron & May \\
 & Blair & Wood \\
 & Brown & Cameron \\
\hline
Salmond & Sturgeon & Sturgeon \\
 & Blair & Wood \\
 & Cameron & Foster \\
\hline
\end{tabular}

\caption{Same as Table \ref{tab8} for leading UK politicians 
and the English Wikipedia edition of 2013.}
\label{tab9}
\end{center}
\end{table}
%%%%%%%%%%%%%%%%%%%%%%%%%%%%%%%%%%%%%%%%%%%%%%%%%%%%%

The ranking index $K_G$ of UK politicians 
from $G_{rr}+G_{\rm qrnd}$ is given in the last column of Table~\ref{tab3}.
It places on top positions Brown, Blair, Cameron
followed by Milibarnd and Salmond. Such a ranking 
looks to be less natural as compared to the global rank index $K$. It 
stresses that that the projector component $G_{\rm pr}$ still plays an 
important role.  

\section{Direct and hidden links of DE politicians}

%%%%%%%%%%%%%%%%%%%%%%%%%%%%%%%%%%%%%%%%%%%%%%%%%%%%%
\begin{figure}
\begin{center}
\includegraphics[width=0.48\textwidth]{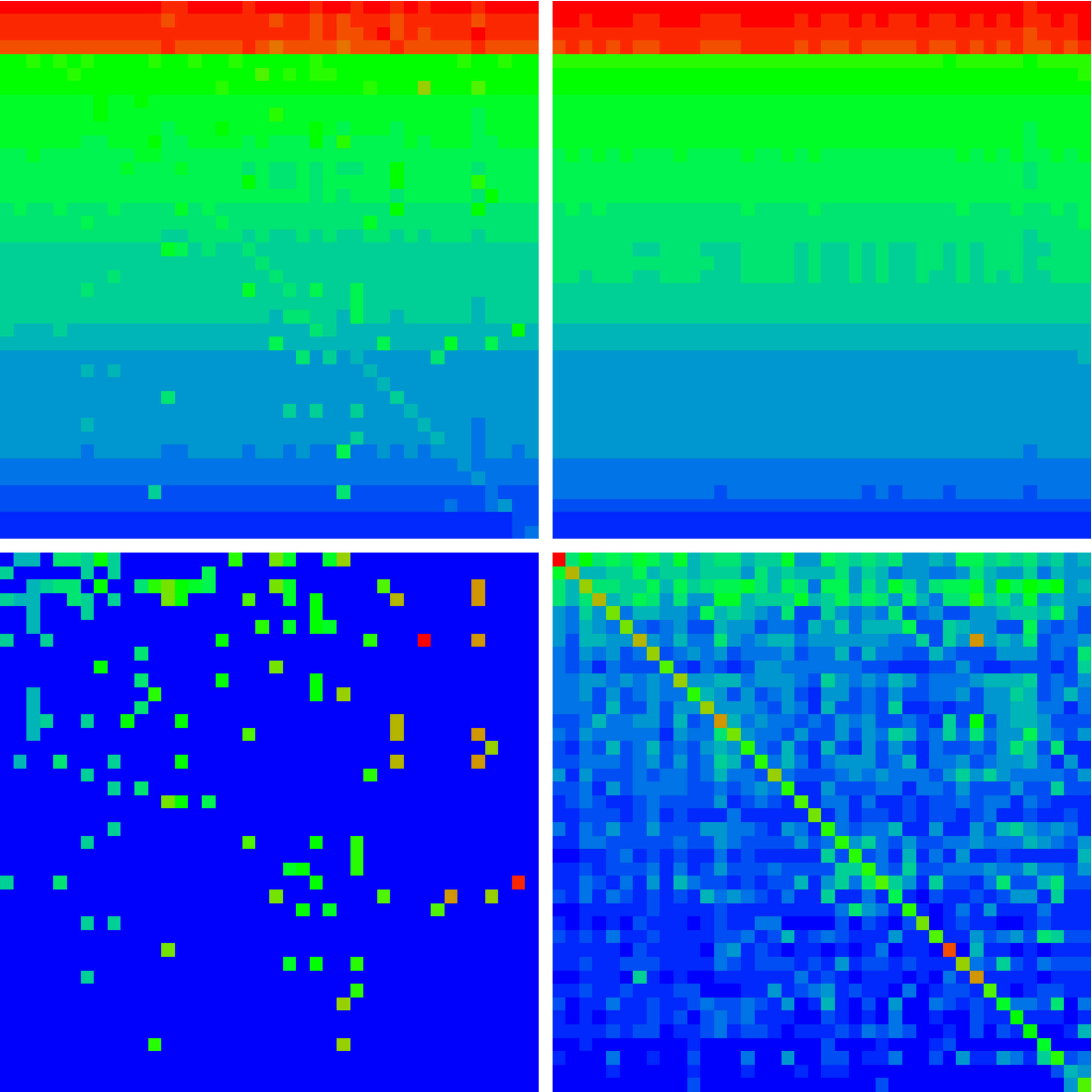}
\caption{Same as Fig. \ref{fig8} for 40 DE politicians in 
the Dewiki network.
The weights of the three matrix components of $G_{\rm R}$ are  
$W_{\rm pr}=0.9590$ for $G_{\rm pr}$, $W_{rr} = 0.02520$ for $G_{rr}$,
$W_{\rm qr}= 0.01580$ for $G_{qr}$. 
}
\label{fig14}
\end{center}
\end{figure}
%%%%%%%%%%%%%%%%%%%%%%%%%%%%%%%%%%%%%%%%%%%%%%%%%%%%%

\begin{figure}
\begin{center}
\includegraphics[width=0.48\textwidth]{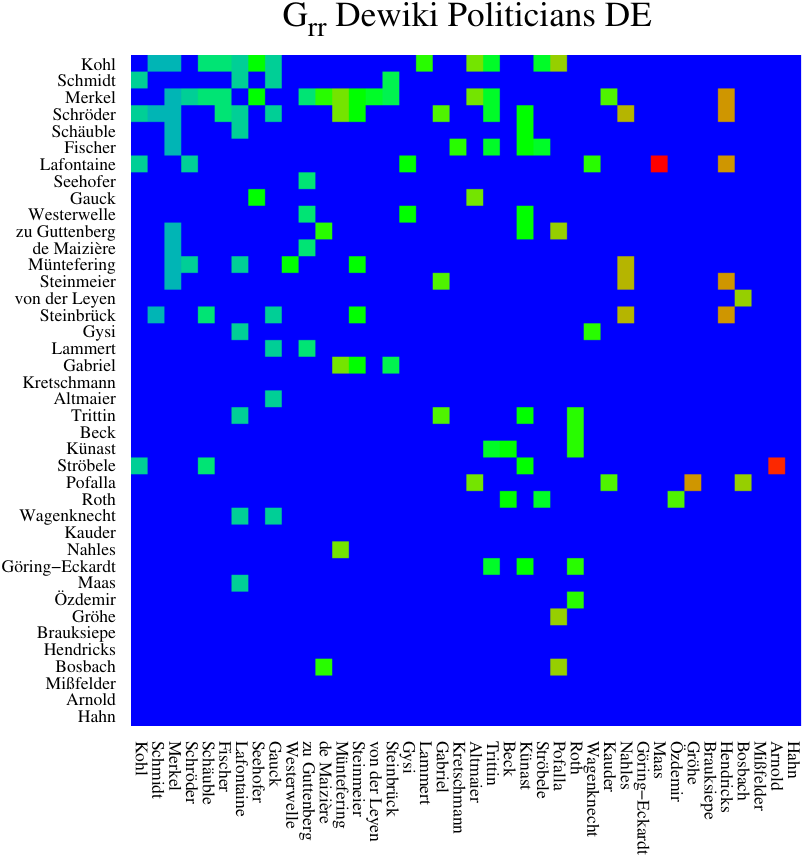}
\caption{Density plot of the matrix $G_{rr}$ for the reduced network 
of 40 DE politicians in the Dewiki network with family names at both axes.
%For a better visibility names for odd and even index numbers are 
%shown in two separate columns. 
}
\label{fig15}
\end{center}
\end{figure}

\begin{figure}
\begin{center}
\includegraphics[width=0.48\textwidth]{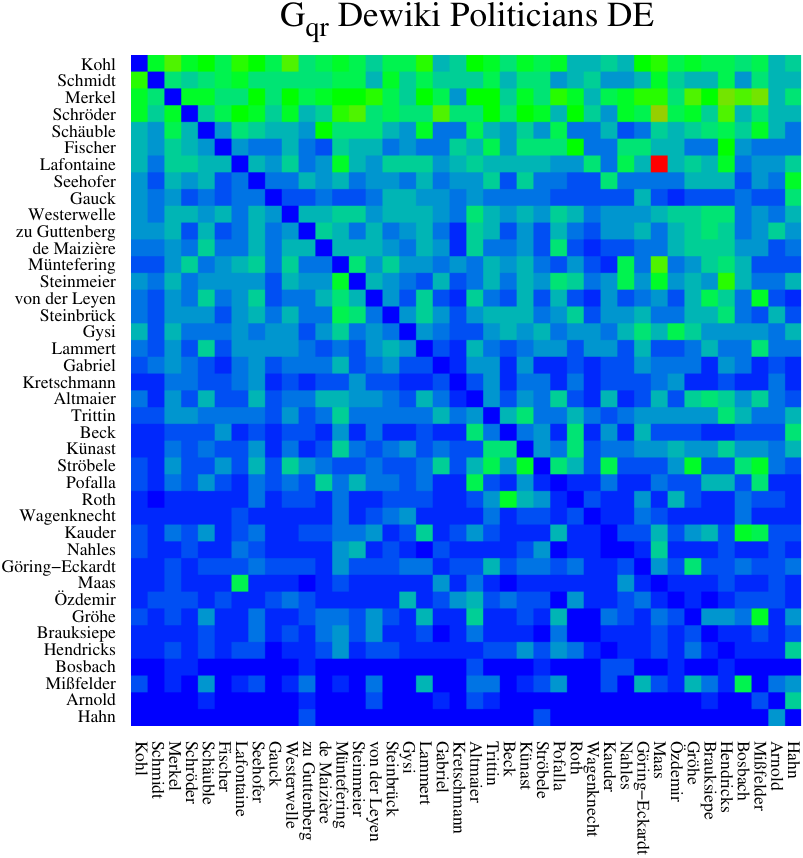}
\caption{Density plot of the matrix $G_{qr}$ without diagonal elements 
for the reduced network of 40 DE politicians in the Dewiki network 
with family names at both axes.
 The weight of this matrix component
without diagonal is $W_{\rm qrnd}= 0.01202$ 
(see Fig.~\ref{fig14} for $W_{\rm qr}$ weight with diagonal).
}
\label{fig16}
\end{center}
\end{figure}

For 40 German politicians the matrix $G_{\rm R}$ and its three components
are shown in Figs.~\ref{fig14},~\ref{fig15} and \ref{fig16}.
The component weights are similar to the cases of US, UK but 
the percentage of $G_{rr}$ is now slightly higher. 

The strongest direct links are from Maas 
to Lafontaine
(Maas was supported by Lafontaine, 
chairman of the Social Democratic Party - SPD)
and Arnold to Str\"obele.

For the indirect links component 
$G_{\rm qr}$ the strongest links remains the one 
of Maas to Lafontaine influenced by their direct link.
However, in global the number of indirect links is 
significantly larger compared to direct links.
The top 5 PageRank politicians 
have strongest direct friend links mainly between their own group 
as it is seen in Table~\ref{tab10}.
However, the list of followers is rather different,
Thus Merkel is the first follower of Kohl who
strongly supported her. Kohl is the strongest follower of 
Schmidt, probably because many Wikipedia articles refer 
to the change of power between them in 1982 but also because Kohl 
was the opposition leader during the Schmidt government of 1974-1982. 
Furthermore, 
Hendricks is the first follower of Merkel. She is a recent member 
of Merkel's government (since end 2013 and at least up to 2016) despite 
being member of the socialist party. The strongest follower 
of Sch\"oder is Maas. Despite the fact that both belong to the socialist party 
it is difficult to establish a direct political link due to a considerable 
time difference of more than 10 years between their time periods in 
government. However, their names appear both (and together with other 
names) in certain news articles discussing general 
marital problems of certain German politicians which may indeed 
produce indirect links on a different than professional level. 

%%%%%%%%%%%%%%%%%%%%%%%%%%%%%%%%%%%%%%%%%%%%%%%%%%%%%
%%%%%%%%%%%%%%%%%%%%%%%%%%%%%%%%%%%%%%%%%%%%%%%%%%%%%

%%%%%%%%%%%%%%%%%%%%%%%%%%%%%%%%%%%%%%%%%%%%%%%%%%%%%
%%%%%%%%%%%%%%%%%%%%%%%%%%%%%%%%%%%%%%%%%%%%%%%%%%%%%

%%%%%%%%%%%%%%%%%%%%%%%%%%%%%%%%%%%%%%%%%%%%%%%%%%%%%
\begin{table}
\begin{center}
\begin{tabular}{|l|l|l|}
\hline
Politicians & DE & Dewiki \\
\hline
\hline
Name & Friends & Followers \\
\hline
\hline
Kohl & Schmidt & Merkel \\
 & Merkel & Westerwelle \\
 & Schröder & Lafontaine \\
\hline
Schmidt & Kohl & Kohl \\
 & Merkel & Lafontaine \\
 & Schröder & Steinbrück \\
\hline
Merkel & Kohl & Hendricks \\
 & Schröder & Mißfelder \\
 & Schäuble & Gröhe \\
\hline
Schröder & Kohl & Maas \\
 & Merkel & Steinmeier \\
 & Schmidt & Hendricks \\
\hline
Schäuble & Kohl & Maizière \\
 & Merkel & Mißfelder \\
 & Schmidt & Lammert \\
\hline
\end{tabular}

\caption{Same as Table \ref{tab8} for leading DE politicians 
and the German Wikipedia edition of 2013.}
\label{tab10}
\end{center}
\end{table}
%%%%%%%%%%%%%%%%%%%%%%%%%%%%%%%%%%%%%%%%%%%%%%%%%%%%%

The ranking $K_G$ from the matrix $G_{rr}+G_{\rm qrnd}$ 
is given for 40 German politicians in the last column
of Table~\ref{tab4} placing on the top positions
Merkel, Schr\"oder, Kohl.

\section{Direct and hidden links of FR politicians}

\begin{figure}
\begin{center}
\includegraphics[width=0.48\textwidth]{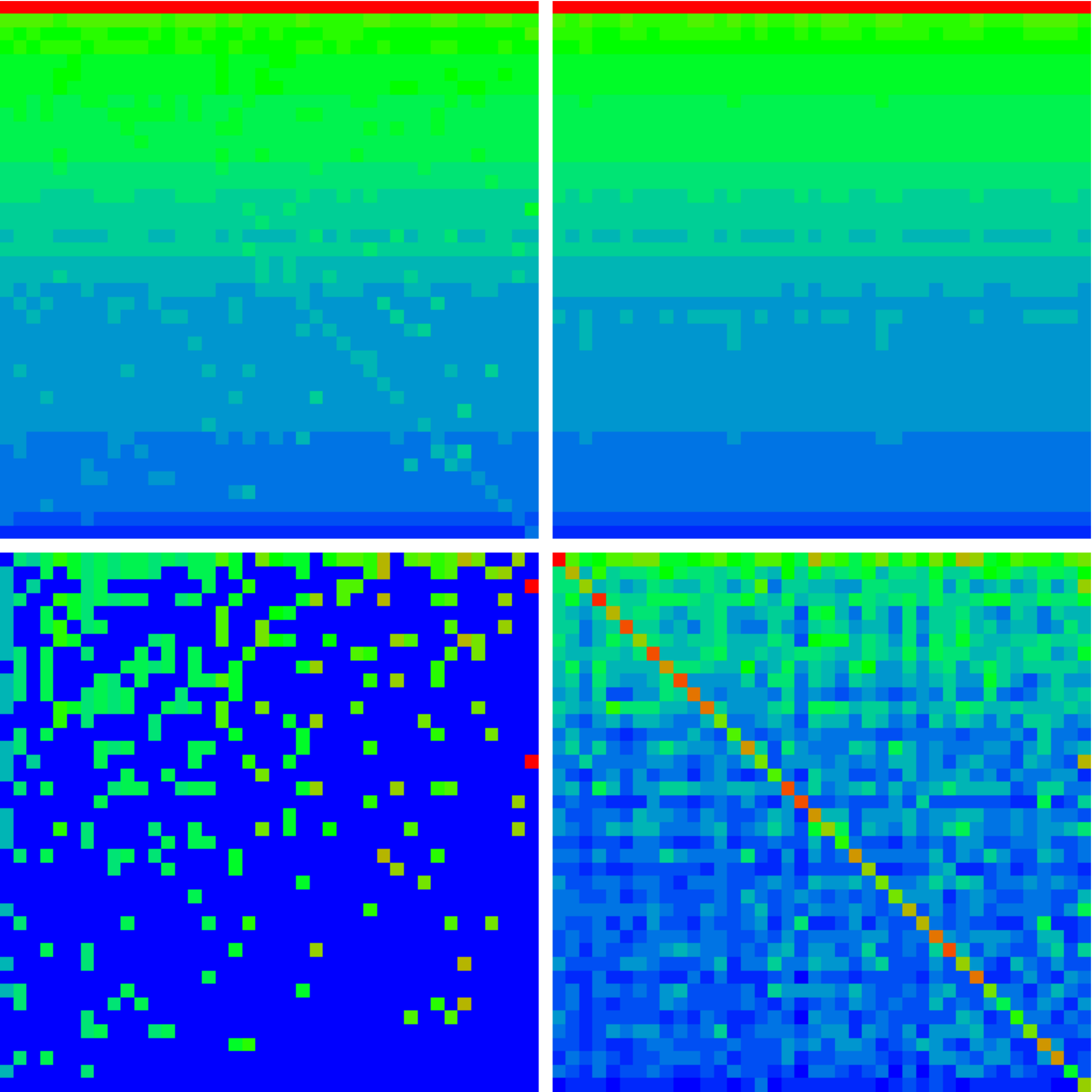}
\caption{Same as Fig. \ref{fig8} for 40 FR politicians in 
the Frwiki network.
The weights of the three matrix components of $G_{\rm R}$ are  
$W_{\rm pr}=0.9481$ for $G_{\rm pr}$, $W_{rr} = 0.03248$ for $G_{rr}$,
$W_{\rm qr}= 0.01941$ for $G_{qr}$. 
}
\label{fig17}
\end{center}
\end{figure}

\begin{figure}
\begin{center}
\includegraphics[width=0.48\textwidth]{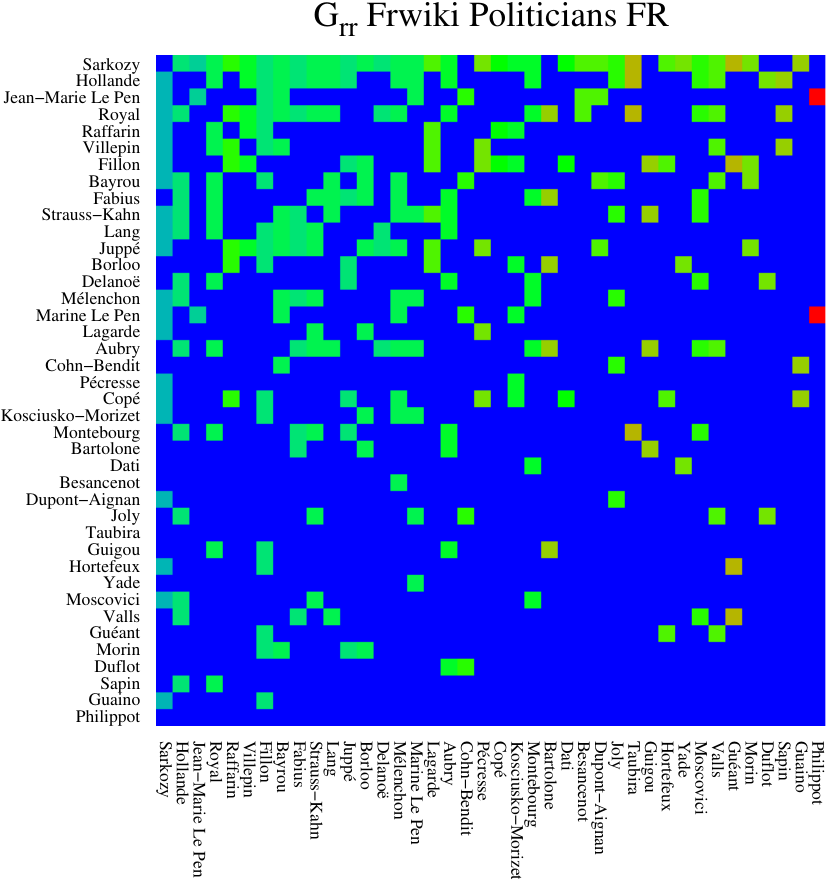}
\caption{Density plot of the matrix $G_{rr}$ for the reduced network 
of 40 FR politicians in the Frwiki network with family names at both axes.
%For a better visibility names for odd and even index numbers are 
%shown in two separate columns. 
}
\label{fig18}
\end{center}
\end{figure}

\begin{figure}
\begin{center}
\includegraphics[width=0.48\textwidth]{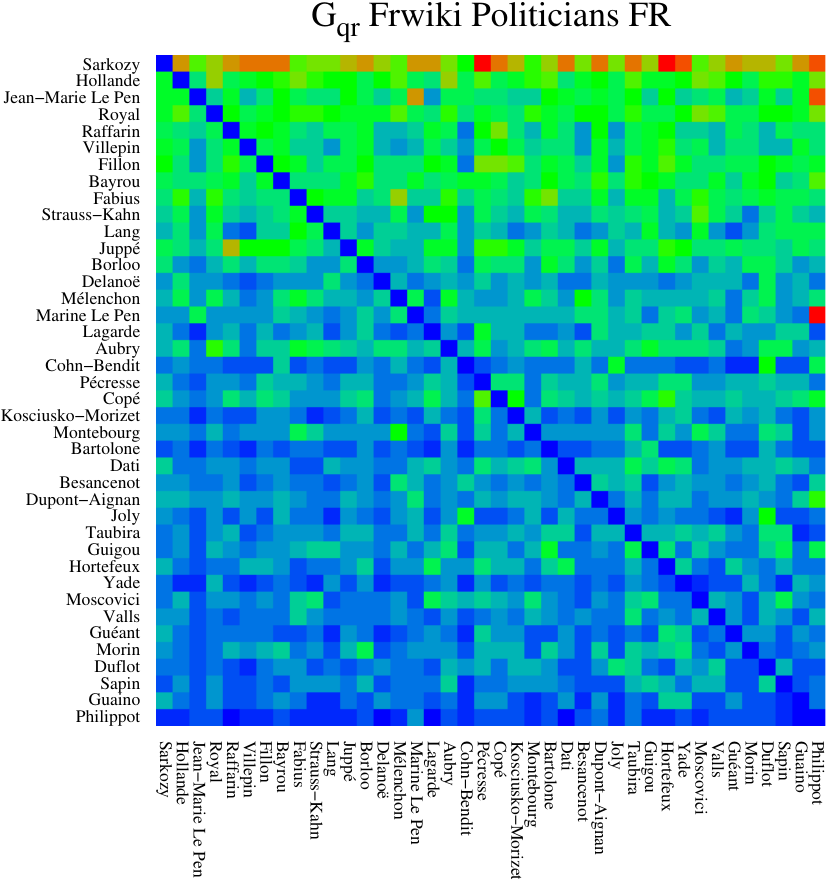}
\caption{Density plot of the matrix $G_{qr}$ without diagonal elements 
for the reduced network of 40 FR politicians in the Frwiki network 
with family names at both axes.
 The weight of this matrix component
without diagonal is $W_{\rm qrnd}= 0.01543$ 
(see Fig.~\ref{fig17} for $W_{\rm qr}$ weight with diagonal).
}
\label{fig19}
\end{center}
\end{figure}

The reduced Google matrix $G_{\rm R}$ and its three
components for 40 French politicians 
(see Fig.~\ref{fig5} and Table~\ref{tab5}) are shown in
Figs.~\ref{fig17},~\ref{fig18},~\ref{fig19}. Here the weight
of $W_{rr}$ is the largest among all groups of politicians 
considered in this work. Also the distribution of direct links 
(see Fig.~\ref{fig18}) is
very broad compared to the case of 40 DE politicians
in Fig.~\ref{fig15}. The two strongest direct links go from 
Philippot to Jean-Marie Le Pen and Marine Le Pen, respectively. 
They all belong to the far-right FN party.

%%%%%%%%%%%%%%%%%%%%%%%%%%%%%%%%%%%%%%%%%%%%%%%%%%%%%
%%%%%%%%%%%%%%%%%%%%%%%%%%%%%%%%%%%%%%%%%%%%%%%%%%%%%

%%%%%%%%%%%%%%%%%%%%%%%%%%%%%%%%%%%%%%%%%%%%%%%%%%%%%
%%%%%%%%%%%%%%%%%%%%%%%%%%%%%%%%%%%%%%%%%%%%%%%%%%%%%

%%%%%%%%%%%%%%%%%%%%%%%%%%%%%%%%%%%%%%%%%%%%%%%%%%%%%
%%%%%%%%%%%%%%%%%%%%%%%%%%%%%%%%%%%%%%%%%%%%%%%%%%%%%

%%%%%%%%%%%%%%%%%%%%%%%%%%%%%%%%%%%%%%%%%%%%%%%%%%%%%
\begin{table}
\begin{center}
\begin{tabular}{|l|l|l|}
\hline
Politicians & FR & Frwiki \\
\hline
\hline
Name & Friends & Followers \\
\hline
\hline
{\color{blue}Sarkozy CB} & {\color{blue}Fillon CB} & {\color{blue}Hortefeux CB} \\
 & {\color{blue-violet}J.-M. Le Pen CV} & {\color{blue}Pécresse CB} \\
 & {\color{magenta}Hollande CM} & {\color{blue}Yade CB} \\
\hline
{\color{magenta}Hollande CM} & {\color{blue}Sarkozy CB} & {\color{magenta}Royal CM} \\
 & {\color{magenta}Royal CM} & {\color{magenta}Aubry CM} \\
 & {\color{magenta}Fabius CM} & {\color{magenta}Fabius CM} \\
\hline
{\color{blue-violet}J.-M. Le Pen CV} & {\color{blue}Sarkozy CB} & {\color{blue-violet}Philippot CV} \\
 & {\color{blue-violet}M. Le Pen CV} & {\color{blue-violet}M. Le Pen CV} \\
 & {\color{blue}Bayrou CB} & {\color{magenta}Taubira CM} \\
\hline
{\color{magenta}Royal CM} & {\color{magenta}Hollande CM} & {\color{magenta}Moscovici CM} \\
 & {\color{blue}Sarkozy CB} & {\color{blue-violet}Philippot CV} \\
 & {\color{magenta}Fabius CM} & {\color{magenta}Hollande CM} \\
\hline
{\color{blue}Raffarin CB} & {\color{blue}Sarkozy CB} & {\color{blue}Copé CB}\\
 & {\color{blue}Juppé CB} & {\color{blue}Pécresse CB} \\
 & {\color{blue}Fillon CB} & {\color{blue}Hortefeux CB} \\
\hline
\end{tabular}

\caption{Same as Table \ref{tab8} for leading FR politicians 
and the French Wikipedia edition of 2013.}
\label{tab11}
\end{center}
\end{table}
%%%%%%%%%%%%%%%%%%%%%%%%%%%%%%%%%%%%%%%%%%%%%%%%%%%%%

The hidden links between 40 FR politicians
without diagonal terms are shown in Fig.~\ref{fig19}.
The 3 strongest friends and followers are given in 
Table~\ref{tab11}.
Friends of current or former presidents Hollande and Sarkozy are main election opponents or close collaborators (Fillon, Fabius). Raffarin, a former prime minister and leader of UMP party, has only same-party hidden links. This could be explained by the fact that he never participated in the last round of any presidential election. Main relationships of Table~\ref{tab11} are reasonable and can be explained. The most surprising connection is the hidden link from Taubira to J.M. Le Pen (there is no direct link). Taubira issued an important law allowing same-sex marriage in 2013, creating virulent opposition from far-right parties that are documented in Wikipedia.  

As pinpointed earlier and in previous works \cite{eomwiki24}, 
using $G_{rr}$ for extracting knowledge on 
followers is not meaningful due to column normalization. A good example can be highlighted here as in $G_{rr}$, top three followers of J.-M. Le Pen (FN party founder) are F. Philippot, N. Dupont-Aignan and A. Montebourg (socialist party), while the leader of FN, Marine Le Pen, only ranks as its 27th follower (due to her increased number of outgoing links compared to others).  
Looking now at the hidden links in Fig.~\ref{fig19} and at the 3 strongest followers of J.M. Le Pen, Marine Le Pen ranks two, just after F. Philippot, the other important figure of that party.  

Several hidden links appear in matrix $G_{\rm qr}$ as well in this network: Philippot to Sarkozy is strong in Fig.~\ref{fig19} and non-existing in Fig.~\ref{fig18} ; same with P\'ecresse to Hollande or Strauss-Kahn to Sarkozy.  
An in-depth analysis of the network of FR politicians is given in Section~\ref{sec:net}.

\section{Direct and hidden links of RU politicians}

\begin{figure}
\begin{center}
\includegraphics[width=0.48\textwidth]{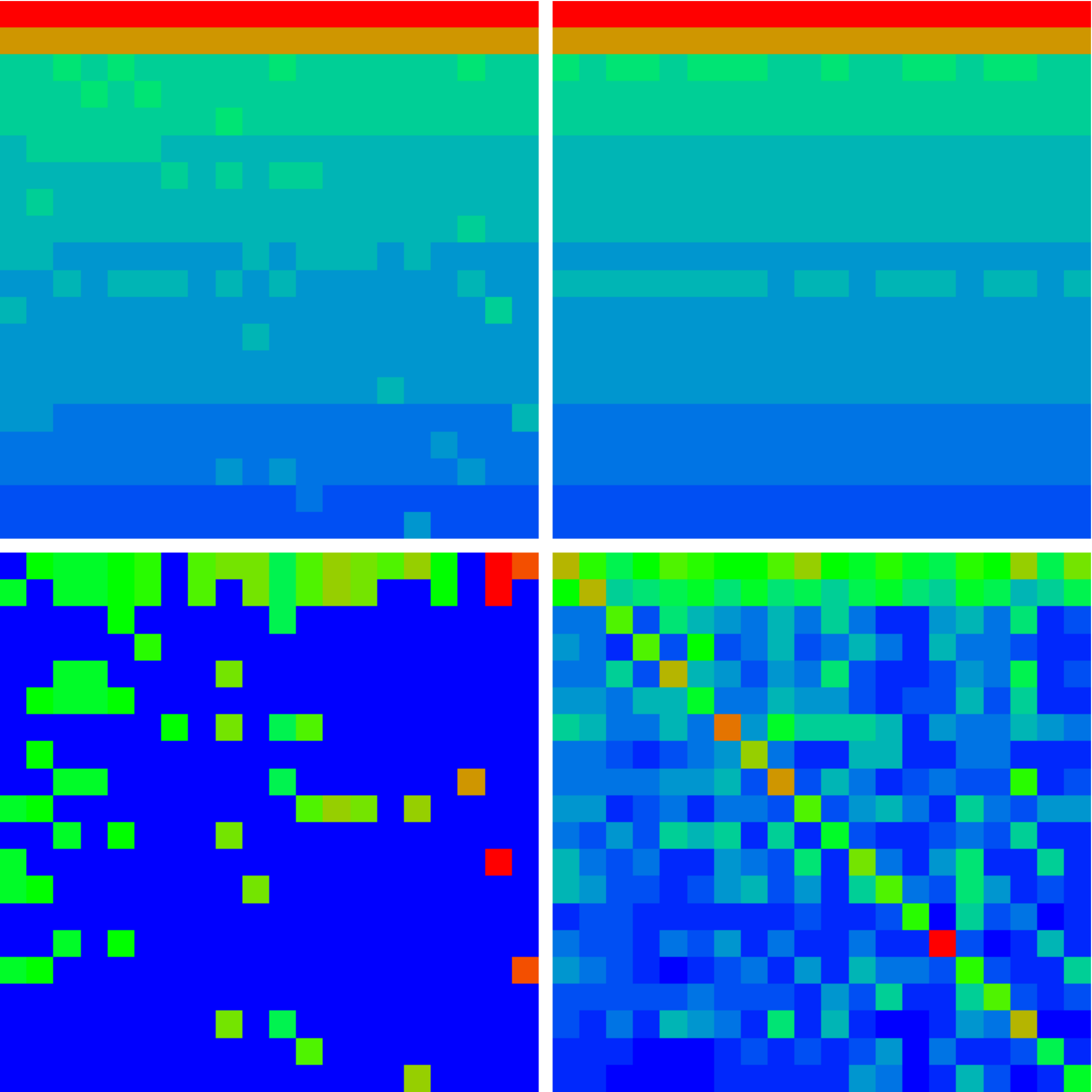}
\caption{Same as Fig. \ref{fig8} for 20 RU politicians in 
the Ruwiki network.
The weights of the three matrix components of $G_{\rm R}$ are  
$W_{\rm pr}=0.9644$ for $G_{\rm pr}$, $W_{rr} = 0.02408$ for $G_{rr}$,
$W_{\rm qr}= 0.011148$ for $G_{qr}$. 
}
\label{fig20}
\end{center}
\end{figure}

\begin{figure}
\begin{center}
\includegraphics[width=0.48\textwidth]{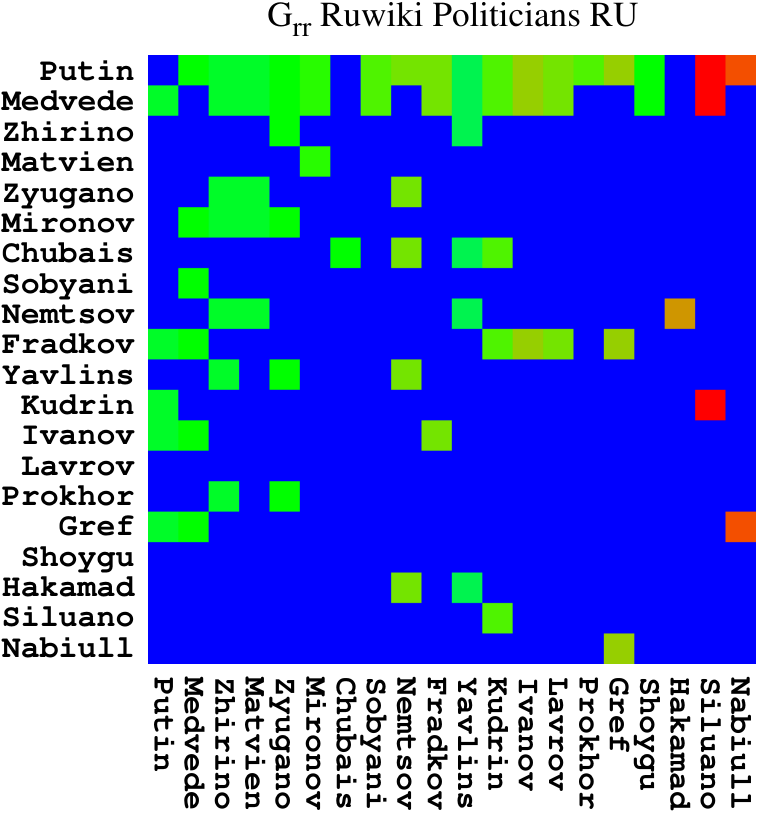}
\caption{Density plot of the matrix $G_{rr}$ for the reduced network 
of 20 RU politicians in the Ruwiki network with short names at both axes.}
\label{fig21}
\end{center}
\end{figure}

\begin{figure}
\begin{center}
\includegraphics[width=0.48\textwidth]{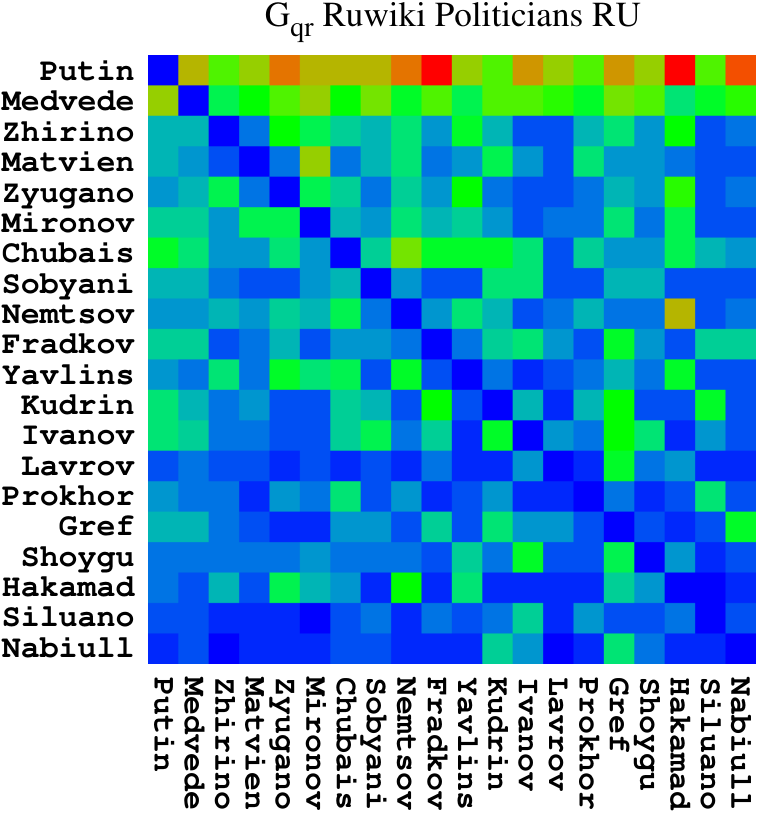}
\caption{Density plot of the matrix $G_{qr}$ without diagonal elements 
for the reduced network of 20 RU politicians in the Ruwiki network 
with short names at both axes.
 The weight of this matrix component
without diagonal is $W_{\rm qrnd}= 0.007912$ 
(see Fig.~\ref{fig20} for $W_{\rm qr}$ weight with diagonal).
}
\label{fig22}
\end{center}
\end{figure}

The reduced Google matrix $G_{\rm R}$ 
for 20 Russian politicians from
Fig.~\ref{fig6} and Table~\ref{tab6} are shown
in Figs.~\ref{fig20},~\ref{fig21},~\ref{fig22}.
The weight of the component of direct links 
is a bit larger than in the case of
US and UK but comparable to DE.
The strongest direct links are from Siluanov (minister of finance)
to Putin and Medvedev and Kudrin (previous minister of finance).

%%%%%%%%%%%%%%%%%%%%%%%%%%%%%%%%%%%%%%%%%%%%%%%%%%%%%
%%%%%%%%%%%%%%%%%%%%%%%%%%%%%%%%%%%%%%%%%%%%%%%%%%%%%

%%%%%%%%%%%%%%%%%%%%%%%%%%%%%%%%%%%%%%%%%%%%%%%%%%%%%
%%%%%%%%%%%%%%%%%%%%%%%%%%%%%%%%%%%%%%%%%%%%%%%%%%%%%

%%%%%%%%%%%%%%%%%%%%%%%%%%%%%%%%%%%%%%%%%%%%%%%%%%%%%
%%%%%%%%%%%%%%%%%%%%%%%%%%%%%%%%%%%%%%%%%%%%%%%%%%%%%

%%%%%%%%%%%%%%%%%%%%%%%%%%%%%%%%%%%%%%%%%%%%%%%%%%%%%
\begin{table}
\begin{center}
\begin{tabular}{|l|l|l|}
\hline
Politicians & RU & Ruwiki \\
\hline
\hline
Name & Friends & Followers \\
\hline
\hline
Putin & Medvedev & Hakamada \\
 & Chubais & Fradkov \\
 & Ivanov & Nabiullina \\
\hline
Medvedev & Putin & Putin \\
 & Chubais & Mironov \\
 & Ivanov & Gref \\
\hline
Zhirinovsky & Putin & Zyuganov \\
 & Medvedev & Hakamada \\
 & Zyuganov & Yavlinsky \\
\hline
Matvienko & Putin & Mironov \\
 & Medvedev & Kudrin \\
 & Mironov & Nemtsov \\
\hline
Zyuganov & Putin & Hakamada \\
 & Medvedev & Yavlinsky \\
 & Zhirinovsky & Zhirinovsky \\
\hline
\end{tabular}

\caption{Same as Table \ref{tab8} for leading RU politicians 
and the Russian Wikipedia edition of 2013.}
\label{tab12}
\end{center}
\end{table}
%%%%%%%%%%%%%%%%%%%%%%%%%%%%%%%%%%%%%%%%%%%%%%%%%%%%%

The matrix of hidden links $G_{\rm qr}$ without diagonal
is shown in Fig.~\ref{fig22} with the list of 2 top friends
and followers for top 5 RU politicians in Table~\ref{tab12}.
Among hidden friends of Putin we find naturally
Medvedev, Chubais, Ivanov who are closely linked
with him during his political career.
More surprising is that his strongest followers are Hakamada, Fradkov, Nabiullina.
Among friend relations it is somewhat unexpected that Zhirinovsky and Zyuganov
are friends of each other in a third position.
It is also interesting that Russian democratically oriented
politicians (Hakamada, Yavlinsky) are among top followers
of Zhirinovsky and Zyuganov.
More naturally, top 2 to 5 PageRank politicians
(Medvedev, Zhirikovsly, Matvienko, Zyuganov)
have Putin as the first friend
(see data at \cite{ourwebpage}).
In fact for all 19 politicians the first friend from the 
matrix of indirect links $G_{\rm qr}$ is Putin
confirming well the vertical power of politics in Russia.
On a level of direct friends from $G_{rr}$ only 
Chubais and Hakamada do not point directly to Putin
but on a level of indirect links of $G_{\rm qr}$
all 19 politicians point first to Putin. 
For RU democrats like we have among top 3 friends (followers)\\
for Yavlinky: Putin, Zyuganov, Chubais\\ 
(Hakamada, Nemtsov, Zyuganov);\\
for Nemtsov: Putin, Chubais, Hakamada \\
(Hakamada, Chubais, Yavlinsky);\\
for Hakamada: Putin, Nemtsov, Zyuganov \\
(Nemtsov, Zyuganov, Yavlinksy).\\
This shows relatively strong links between democrats even
contacts with Zyuganov (leader of communist party)
are surprisingly well present. 

The rank index $K_G$ from the matrix $G_{rr}+G_{\rm qrnd}$
is given in last column of Table~\ref{tab6}
with the top leaders being Putin, Medvedev, Fradkov that seems to 
overestimate the importance of Fradkov, even if he is followed by 
Ivanov and Gref that looks to be more reasonable.
So the above analysis of politicians of US, UK, DE, FR, RU
suggests that the projector component 
should be taken into account even if we want to analyze the relations
inside the selected group of politicians since the environment
links of the global matrix $G$ still play an important role. 

\section{Direct and hidden links of G20 state leaders}

Above we considered interactions between political leaders of 
the same country
from the view point of the Wikipedia edition of their main language. 
It is interesting to see 
the results of the reduced Google matrix analysis
for interactions of state leaders of the G20 summit of 2012 \cite{g20wiki}.
We analyze these interactions from the view point
of 4 Wikipedia editions EN, DE, FR, RU.

%%%%%%%%%%%%%%%%%%%%%%%%%%%%%%%%%%%%%%%%%%%%%%%%%%%%%
\begin{figure}
\begin{center}
\includegraphics[width=0.48\textwidth]{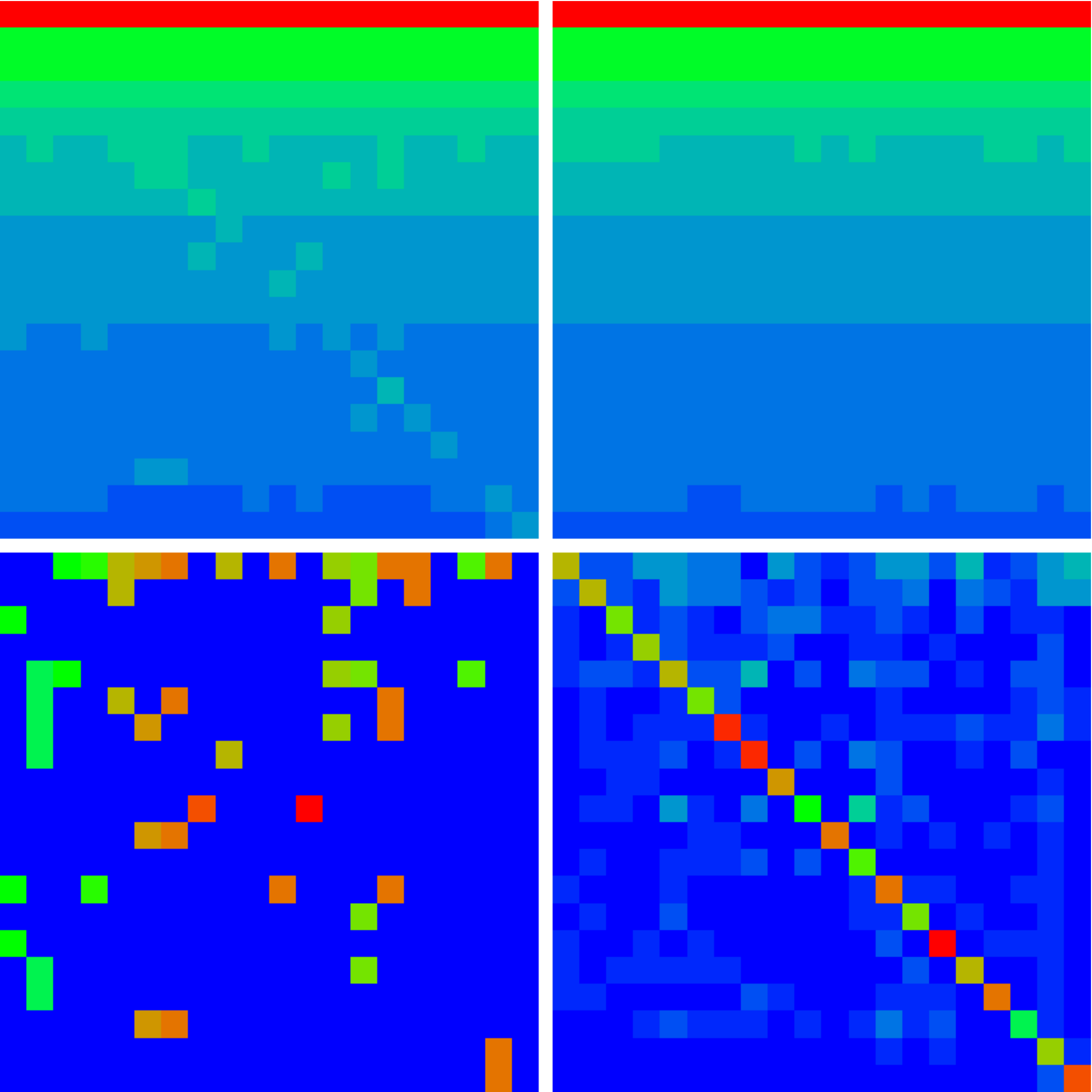}
\caption{Same as Fig. \ref{fig8} for 20 state leaders of G20 states 
in the Enwiki network.
The weights of the three matrix components of $G_{\rm R}$ are  
$W_{\rm pr}=0.9793$ for $G_{\rm pr}$, $W_{rr} = 0.009976$ for $G_{rr}$,
$W_{\rm qr}= 0.01070$ for $G_{qr}$. 
}
\label{fig23}
\end{center}
\end{figure}
%%%%%%%%%%%%%%%%%%%%%%%%%%%%%%%%%%%%%%%%%%%%%%%%%%%%%

%%%%%%%%%%%%%%%%%%%%%%%%%%%%%%%%%%%%%%%%%%%%%%%%%%%%%
\begin{figure}
\begin{center}
\includegraphics[width=0.48\textwidth]{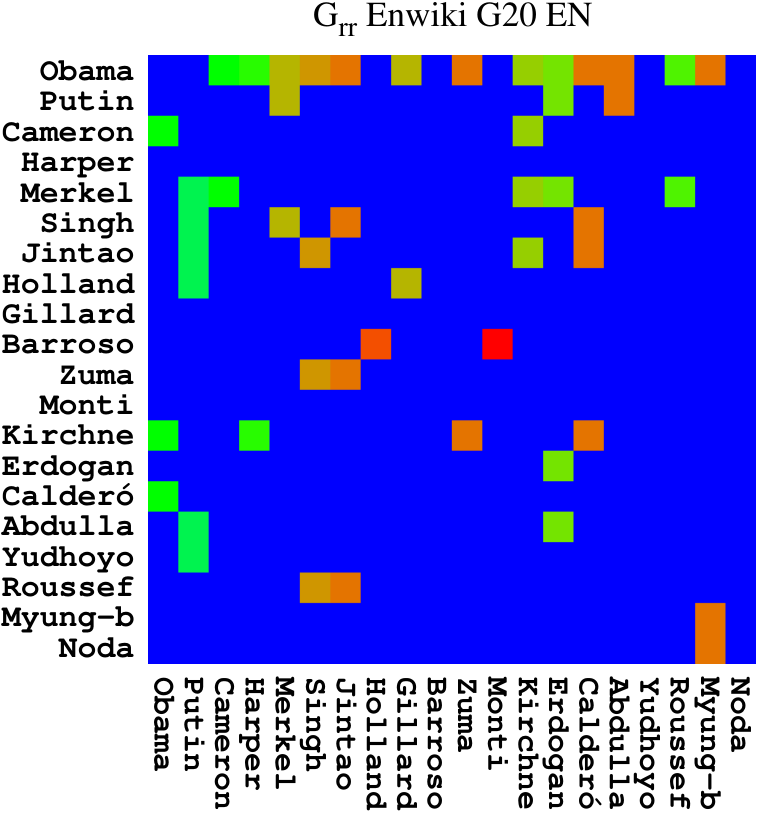}
\caption{Density plot of the matrix $G_{rr}$ for the reduced network 
of 20 state leaders of G20 states in the Enwiki network with 
short names at both axes.}
\label{fig24}
\end{center}
\end{figure}
%%%%%%%%%%%%%%%%%%%%%%%%%%%%%%%%%%%%%%%%%%%%%%%%%%%%%

\begin{figure}
\begin{center}
\includegraphics[width=0.48\textwidth]{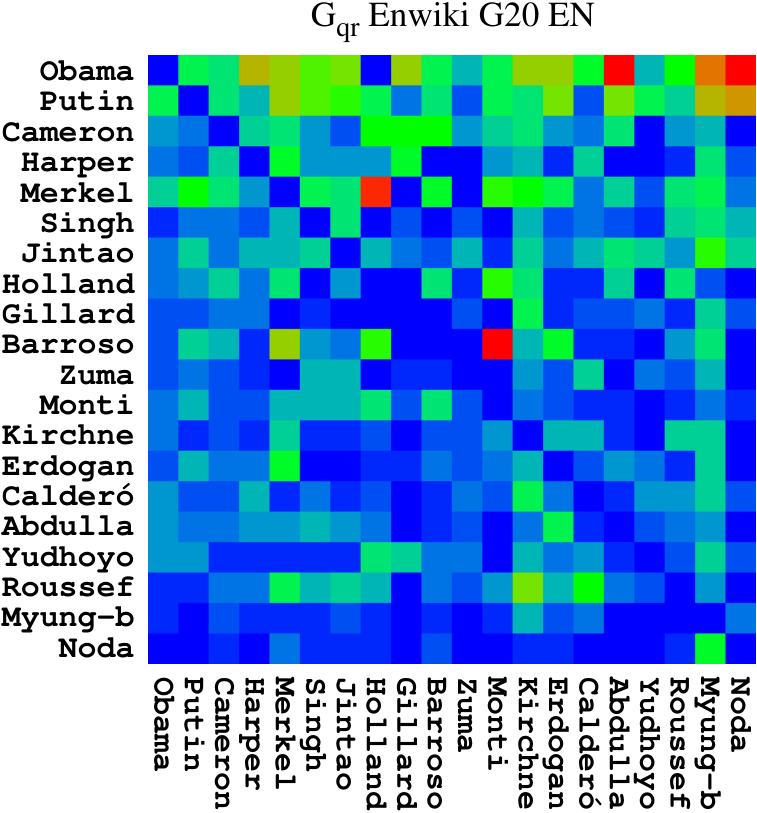}
\caption{Density plot of the matrix $G_{qr}$ without diagonal elements 
for the reduced network 
of 20 state leaders of G20 states in the Enwiki network with 
short names at both axes.
 The weight of this matrix component
without diagonal is $W_{\rm qrnd}= 0.001878$ 
(see Fig.~\ref{fig23} for $W_{\rm qr}$ weight with diagonal).
}
\label{fig25}
\end{center}
\end{figure}

The list of names of politicians of G20 and their
distribution over PageRank-CheiRank plane are given in
Fig.~\ref{fig7} and Table~\ref{tab7}.
In our presentation below we keep the names in the PageRank order
of Enwiki of Table~\ref{tab7}.
For EN Wikipedia the reduced Google matrix $G_{\rm R}$
and its three components are shown in 
Figs.~\ref{fig23},~\ref{fig24},~\ref{fig25}.
We see that, compared to previous cases inside one country, 
the weights $W_{rr}$ and $W_{\rm qr}$
are  reduced approximately by a factor 2.
Indeed, there are significantly less direct links
between leaders of different states (see Fig.~\ref{fig24}).
There are even less direct links for DE, FR, RU editions
(see data at \cite{ourwebpage}). For example, the only direct links
for RU Wikipedia are between Merkel and Putin, Erdoğan,
Putin and Jintao. 

Thus the importance of indirect links from $G_{\rm qr}$
becomes more significant even if the weight of
nondiagonal matrix elements is reduced by a factor 4-5
compared to the case of politicians in the same country.
In Fig.~\ref{fig25} for Enwiki we find the strongest indirect links
between Monti and Barroso (EU), Abdullah (SA) and Obama
(even if direct links also exist), Hollande and Merkel,
Noda (JP) and Obama (without direct links).
The list of top 3 friends (followers) of top 5 PageRank leaders
of Enwiki is given in Table~\ref{tab13}. Among friends
the first positions are taken by Putin, Merkel, Obama.
For Merkel the top friends are Barroso, Putin, Obama;
for Cameron we have Putin, Obama, Merkel.
Among top followers of Obama we do not find EU leaders
but Noda (JP), Abdullah (SA), Myung-bak (KR). The followers of Putin are
Noda, Mying-bak, Merkel.
The first follower of Merkel is Hollande while Merkel is the top follower of
Harper (CA).

%%%%%%%%%%%%%%%%%%%%%%%%%%%%%%%%%%%%%%%%%%%%%%%%%%%%%
%%%%%%%%%%%%%%%%%%%%%%%%%%%%%%%%%%%%%%%%%%%%%%%%%%%%%

%%%%%%%%%%%%%%%%%%%%%%%%%%%%%%%%%%%%%%%%%%%%%%%%%%%%%
\begin{table}
\begin{center}
\begin{tabular}{|l|l|l|}
\hline
G20 & EN & Enwiki \\
\hline
\hline
Name & Friends & Followers \\
\hline
\hline
Obama & Putin & Noda \\
 & Merkel & Abdullah \\
 & Calderón & Myung-bak \\
\hline
Putin & Merkel & Noda \\
 & Obama & Myung-bak \\
 & Barroso & Merkel \\
\hline
Cameron & Putin & Gillard \\
 & Obama & Barroso \\
 & Merkel & Hollande \\
\hline
Harper & Obama & Merkel \\
 & Cameron & Gillard \\
 & Putin & Myung-bak \\
\hline
Merkel & Barroso & Hollande \\
 & Putin & Monti \\
 & Obama & Kirchner \\
\hline
\end{tabular}

\caption{Same as Table \ref{tab8} for leading G20 state leader 
and the English Wikipedia edition of 2013.}
\label{tab13}
\end{center}
\end{table}
%%%%%%%%%%%%%%%%%%%%%%%%%%%%%%%%%%%%%%%%%%%%%%%%%%%%%

\begin{figure}
\begin{center}
\includegraphics[width=0.48\textwidth]{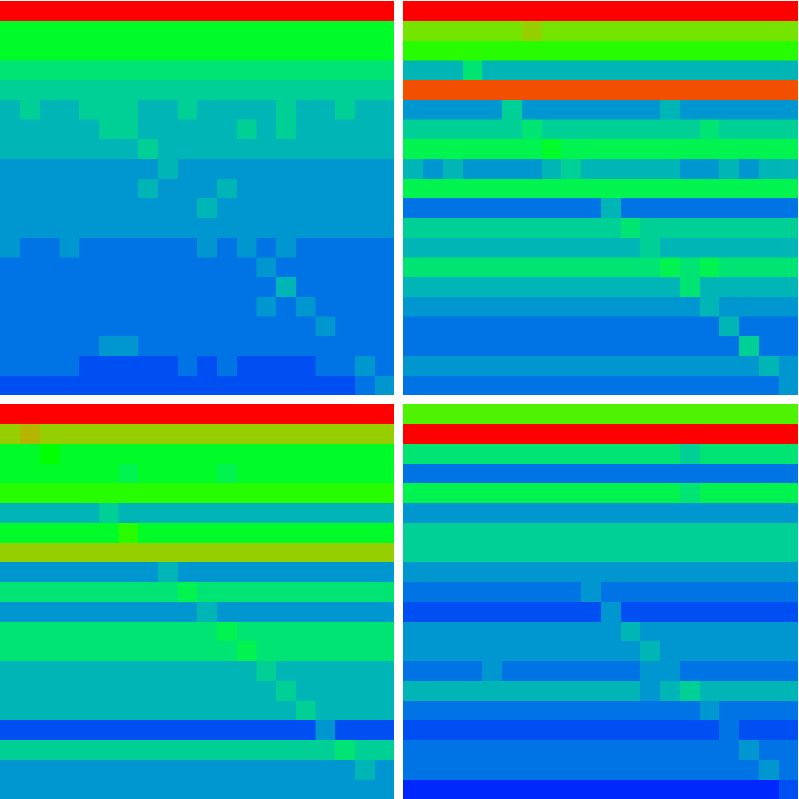}
\caption{Density plots of the matrix $G_{\rm R}$ 
for the reduced networks of 20 state leaders of G20 states 
in the Enwiki (top left), Dewiki (top right), Frwiki (bottom left) 
and Ruwiki (bottom right) networks. The order of the 20 state leaders 
is in all cases given by the PageRank order for the Enwiki network 
and corresponds to the same order of Figs. \ref{fig23}-\ref{fig25}.}
\label{fig26}
\end{center}
\end{figure}

The reduced Google matrices $G_{\rm R}$ 
from 4 Wikipedia editions are shown in Fig.~\ref{fig26}.
It is clear that each culture 
(which, in first approximation, 
can be associated with the language) has its own
view on relations between state leaders of G20.
Indeed, even top 3 PageRank leaders are different
for these cultures which creates different structures of 
matrix elements.  

%%%%%%%%%%%%%%%%%%%%%%%%%%%%%%%%%%%%%%%%%%%%%%%%%%%%%
%%%%%%%%%%%%%%%%%%%%%%%%%%%%%%%%%%%%%%%%%%%%%%%%%%%%%

\begin{figure}
\begin{center}
\includegraphics[width=0.48\textwidth]{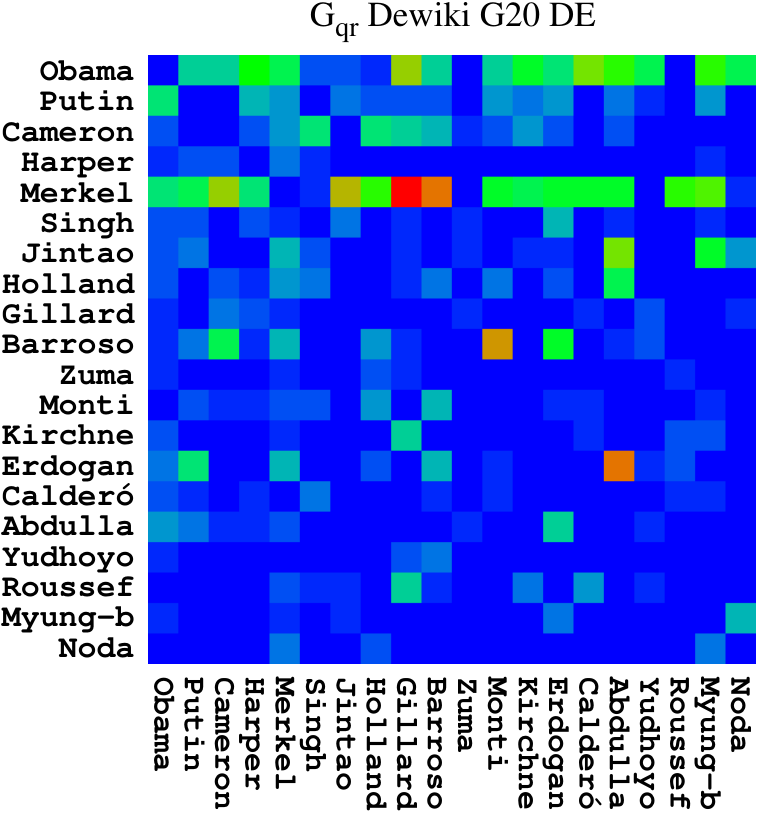}
\caption{Density plot of the matrix $G_{qr}$ without diagonal elements 
for the reduced network of 20 state leaders of G20 states in the Dewiki 
network with short names (in English) at both axes using the same order of 
Fig. \ref{fig25}.
 The weight of this matrix component
without diagonal is $W_{\rm qrnd}= 0.001125$. 
}
\label{fig27}
\end{center}
\end{figure}

\begin{figure}
\begin{center}
\includegraphics[width=0.48\textwidth]{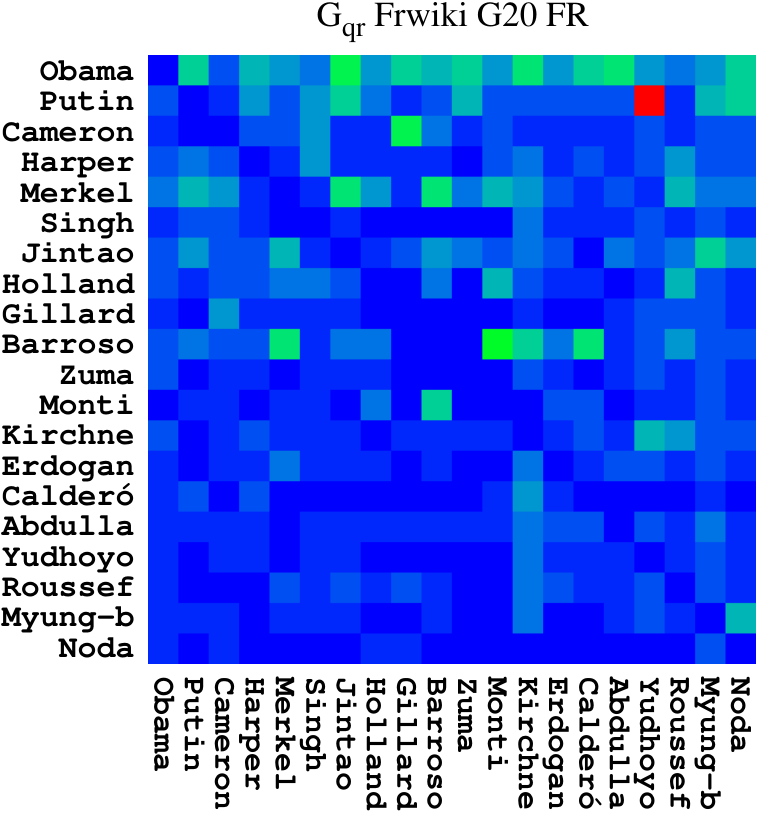}
\caption{Density plot of the matrix $G_{qr}$ without diagonal elements 
for the reduced network of 20 state leaders of G20 states in the Frwiki 
network with short names (in English) at both axes using the same order of 
Fig. \ref{fig25}.
 The weight of this matrix component
without diagonal is $W_{\rm qrnd}= 0.001554$. 
%The norm number is $N_G=0.001554$.
}
\label{fig28}
\end{center}
\end{figure}

\begin{figure}
\begin{center}
\includegraphics[width=0.48\textwidth]{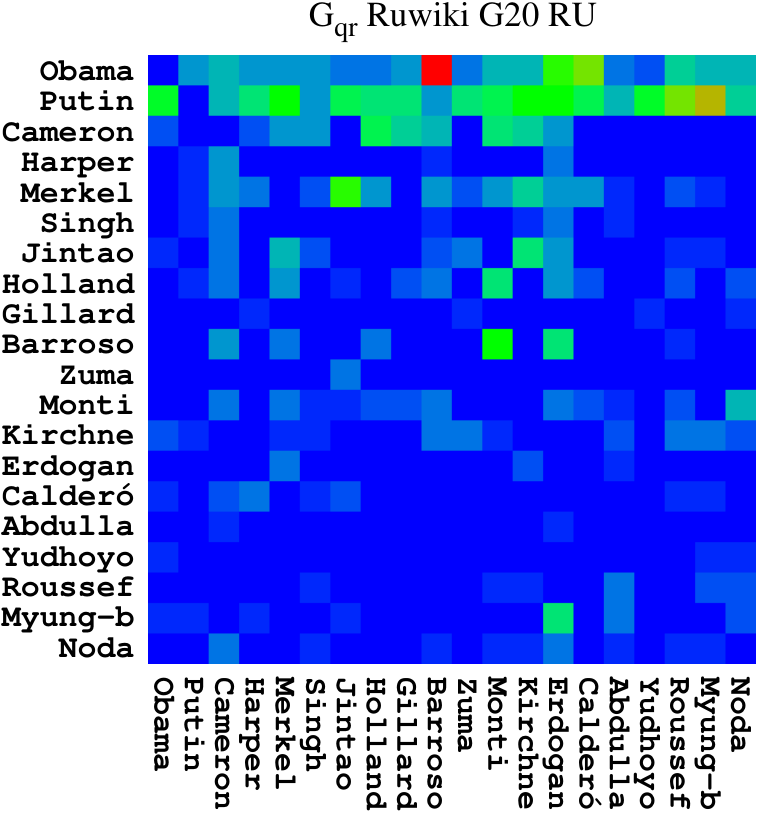}
\caption{Density plot of the matrix $G_{qr}$ without diagonal elements 
for the reduced network of 20 state leaders of G20 states in the Ruwiki 
network with short names (in English) at both axes using the same order of 
Fig. \ref{fig25}.
 The weight of this matrix component
without diagonal is $W_{\rm qrnd}= 0.001089$. 
%The norm number is $N_G=0.001089$.
}
\label{fig29}
\end{center}
\end{figure}

The indirect links of $G_{\rm qr}$ are shown in Fig.~\ref{fig27}
for Dewiki, Fig.~\ref{fig28} for Frwiki
and Fig.~\ref{fig29} for Ruwiki. 
For Dewiki the strongest indirect links are
from Gillard (AU) and Barroso to Merkel,
Abdullah to Erdoğan. For Frwiki the strongest link is 
from Yudhoyono (ID) to Putin, Monti and Merkel to Barroso.
In contrast for Ruwiki the strongest links are
from Barroso to Obama, Myung-bak (KR) to Putin.
Being at the top of PageRank in Ruwiki,
Putin accumulates the largest number of followers.
This demonstrates a large variety of cultural views on
interactions between state leaders.

The English version provides the richest information thanks 
to its volume and the variety of its contributions. However, main trends in other countries 
are not necessarily pictured in the English version due to cultural bias, and vice verse. 
For instance, the very strong link from Hollande to Merkel in Enwiki 
is really thin in the French edition, while it is clearly visible in the German one.

%%%%%%%%%%%%%%%%%%%%%%%%%%%%%%%%%%%%%%%%%%%%%%%%%%%%%
%%%%%%%%%%%%%%%%%%%%%%%%%%%%%%%%%%%%%%%%%%%%%%%%%%%%%

%%%%%%%%%%%%%%%%%%%%%%%%%%%%%%%%%%%%%%%%%%%%%%%%%%%%%
%%%%%%%%%%%%%%%%%%%%%%%%%%%%%%%%%%%%%%%%%%%%%%%%%%%%%

%%%%%%%%%%%%%%%%%%%%%%%%%%%%%%%%%%%%%%%%%%%%%%%%%%%%%
%%%%%%%%%%%%%%%%%%%%%%%%%%%%%%%%%%%%%%%%%%%%%%%%%%%%%

\section{Network of political leaders of France}\label{sec:net}

\begin{figure}
\begin{center}
\includegraphics[width=0.48\textwidth]{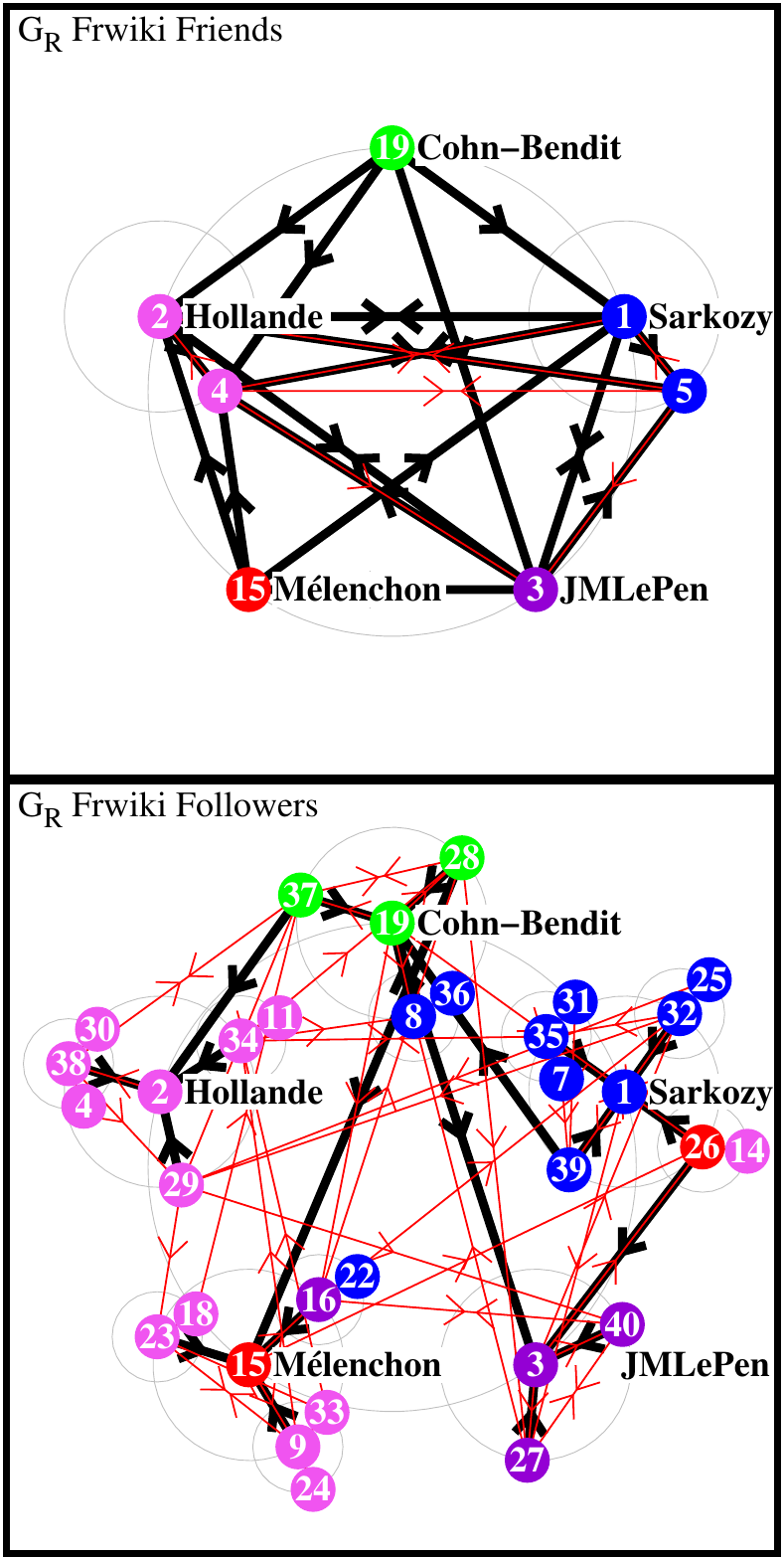}
\caption{Network structure of friends (followers) in 
top (bottom) panel induced by 
the largest four nondiagonal elements per column (row) of the 
matrix $G_{\rm R}$ obtained for the group of 40 French politicians in the 
French Wikipedia edition of 2013. The five colors of the nodes represent 
the five main political movements and the numbers of the nodes represent 
the local PageRank index in Table~\ref{tab5}. 
For each movement the leading node in PageRank 
ordering is drawn on the largest primary circle with name. 
The thick black arrows represent 
links from (to) each primary node to (from) its four most relevant nodes and 
the thin red arrows represent links from (to) each secondary node to (from) 
its four most relevant nodes. New secondary (tertiary) nodes are drawn on the 
secondary (tertiary) circles. Most of these circles contain less than 
four nodes due to possible multiple links with center nodes 
on other circles of the same level. 
When possible, a secondary (tertiary) node was associated 
to a primary (secondary) node of the same color. 
}
\label{fig30}
\end{center}
\end{figure}

To have a more direct pictorial representation of interactions 
in the framework of the reduced matrix $G_{\rm R}$ we choose the case of 
40 politicians of France discussed above (see Fig.~\ref{fig5}
and Table~\ref{tab5}). Here all politicians are attributed 
to the main political parties marked by corresponding 5 colors
in Table~\ref{tab5}. For each color we take the top
PageRank politicians:
Sarkozy (blue, right parties, UMP-UDI),
Hollande (magenta, left, PS), Jean-Marie Le Pen
(violet, ultra-right, FN), M\'elanchon (red, ultra-left),
Cohn-Bendit (green, green party). These 5 ``leaders''\footnote{
This definition of leader does not necessarily correspond 
to the official ``leader'' of a political party or group since 
in the Wikipedia PageRank index also older, more historical aspects, e.~g. 
``former leadership'', are taken into account. }
are positioned on a circle 
of fixed radius forming the first level.
Then for the directed network 
of $G_{\rm R}$ of these 5 leaders
we show for each of them 4 strongest 
links from them to other politicians (considered as their 
direct friends). These links are shown by directed bold black lines with arrows
in the top panel of Fig.~\ref{fig30}. The new (secondary) 
politicians appeared from these links are placed on second level
circles around the (primary) politicians of the first level to which they 
correspond (the preference is given to the primary politicians with the same 
color if there are several corresponding primary politicians). 
For the politicians of the second level
new red links with arrows are drawn for each of them with top 4 
strongest links forming the third level circles if any.
After these two iterations of 4 strongest friends we obtain the
network of friends of $G_{\rm R}$ with only 7 politicians.
The politicians are marked by their PageRank numbers $K$ from Table~\ref{tab5}.
At blue color we find Sarkozy with Raffarin, at magenta we have Hollande
with Royal, and for other colors we have only 
one politician of the first circle.
Thus we find that the circle of close friends is very narrow. We also 
mention that for this case the subnetwork actually saturates completely 
at level 3 (i.e. including the group of tertiary politicians visible 
in Fig.~\ref{fig30}) with only these 7 politicians and does not increase 
even if we try to include higher level circles/politicians. 
The reason of the saturation on a small sub network is that $G_{\rm R}$ 
is dominated by the rank one contribution $G_{\rm pr}$ which selects 
essentially top PageRank nodes. Therefore we simply find the top 5 PageRank 
nodes plus the two late PageRank position nodes of M\'elanchon and Cohn-Bendit that have been selected to belong to the set of primary nodes. 

We now perform the same procedure for the followers in $G_{\rm R}$,  
using the strongest 4 incoming links on each level, instead of friends,
as it is shown in the bottom panel of Fig.~\ref{fig30}. 
Again after two iterations
(black and red links) we obtain 32 politicians showing 
that the number of followers
is significantly larger then the number of friends.
The network of followers has an interesting structure:
violet, blue and green parties form mainly compact groups
while the magenta party is clearly divided on two
groups one centered around Hollande and another centered around 
M\'elanchon. This clearly indicates fundamental structural issues
in French PS. We note that for this case of followers (for $G_{\rm R}$) 
the subnetwork would saturate at all 40 politicians 
after 5 iterations (the corresponding higher level links are not shown in 
Fig.~\ref{fig30} in order to keep the presentation simpler). 

%%%%%%%%%%%%%%%%%%%%%%%%%%%%%%%%%%%%%%%%%%%%%%%%%%%%%
%%%%%%%%%%%%%%%%%%%%%%%%%%%%%%%%%%%%%%%%%%%%%%%%%%%%%

%%%%%%%%%%%%%%%%%%%%%%%%%%%%%%%%%%%%%%%%%%%%%%%%%%%%%
\begin{figure}
\begin{center}
\includegraphics[width=0.48\textwidth]{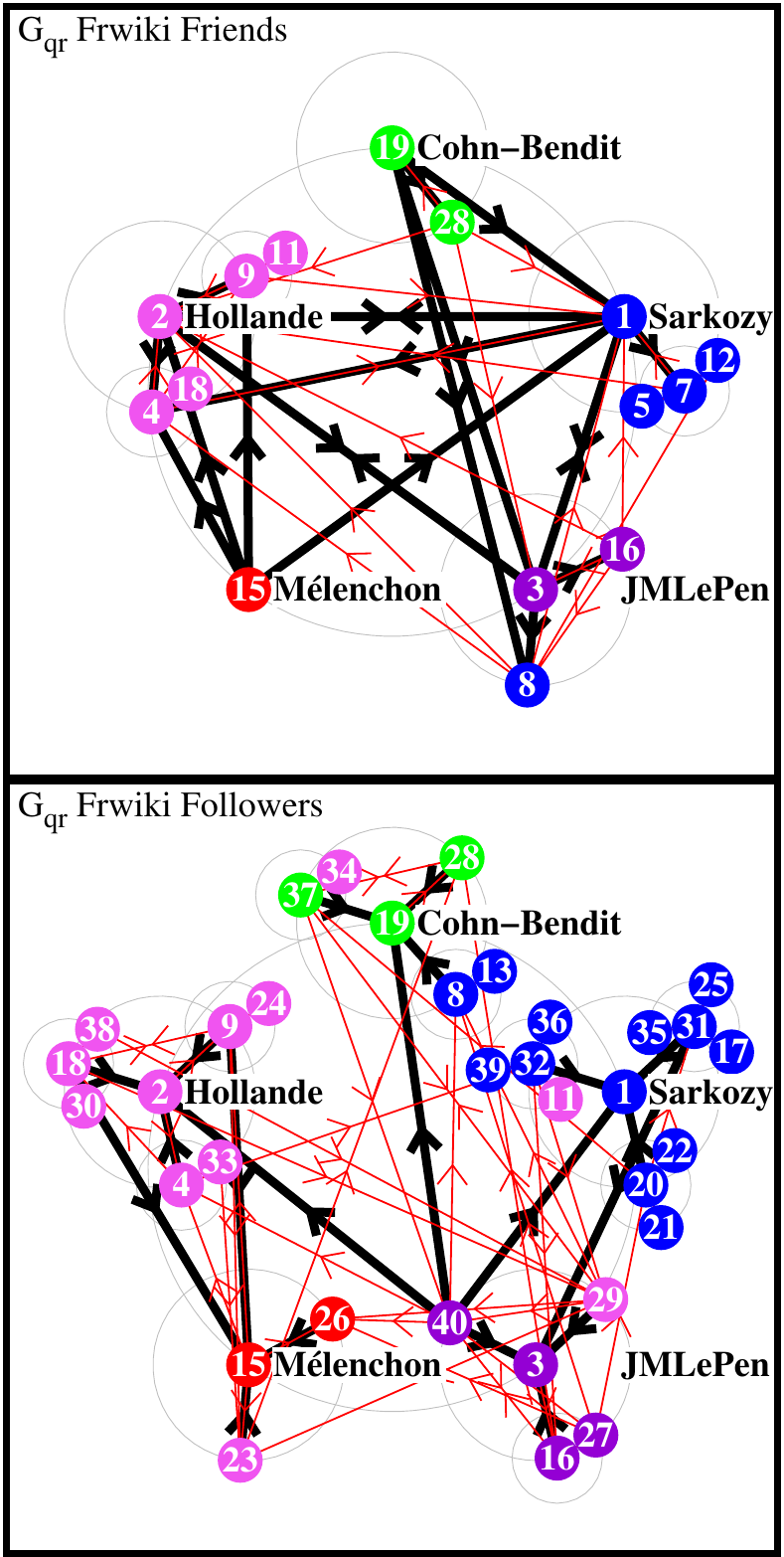}
\caption{Same as Fig.~\ref{fig30} for the matrix $G_{qr}$ 
which recovers hidden links between politicians. 
}
\label{fig31}
\end{center}
\end{figure}
%%%%%%%%%%%%%%%%%%%%%%%%%%%%%%%%%%%%%%%%%%%%%%%%%%%%%

We also used the same approach to construct the network
of hidden friends or hidden followers using the matrix $G_{\rm qr}$ 
(instead of $G_{\rm R}$) with the result shown in both panels of 
Fig.~\ref{fig31}. Now, the network of hidden friends 
contains a significantly larger number of politicians after two iterations
(15 instead of 7 for $G_{\rm R}$ but also with saturation at level 3). 
At the same time the network
of followers has a similar number of nodes 
(34 instead of 32 for $G_{\rm R}$ with saturation at 35 after level 4). 
For the hidden follower network 
the separation of the magenta PS party in two groups
is less pronounced with a smaller number of followers for M\'elanchon.
The group of the violet party becomes more compact
having all its 4 members being grouped together. There is also
Taubira ($K=29$) who is closely linked to this violet
group for similar reasons as the ones presented earlier. 

Several follower interactions seem more reasonable with $G_{\rm qr}$. 
For instance, Duflot ($K=37$, Green party) isn't linked anymore to 
PS but mainly to its own party members. Similarly, ultra-left members 
are connected together at the first level, and not only at 
the secondary level as it is the case for $G_{\rm R}$. 
Interesting to notice is that in $G_{\rm qr}$ Philippot ($K=40$, ultra-right FN party) 
clearly follows four of the main party leaders (J.-M. Le Pen, Sarkozy, 
Hollande and Cohn-Bendit), while this is not the case in $G_{\rm R}$. 
This is clearly related to his political acquaintances of before 
2009 where he backed up 
Chev\`enement (socialist party) and M\'elenchon, among others.   

\section{Discussion}
\label{sec5}

In conclusion, we have presented a new mathematical
meth\-od which establishes an effective directed network
for a selected subset of nodes belonging to a significantly larger network.
This approach was tested on examples of several groups
of political leaders of 5 countries and 
world state leaders of G20 analyzed in the frame of 
several Wikipedia networks. Our results show that the proposed method
allows in a reliable way to determine direct and hidden 
links between political leaders. We think that this approach can provide
firm mathematical grounds for the LMX studies \cite{lmx1,lmx2,lmx3}
in social and political sciences.
Our results show that the Wikipedia network
can be used in an efficient way to determine
direct and hidden relations between different
subjects appearing in Wikipedia. We also point 
that the reduced Google matrix approach
can be applied to a variety of directed networks
where the relations between selected subgroup of nodes 
are not straightforward to identify.

We thank Leonardo Ermann for stimulating discussions
of properties of the reduced Google matrix for various directed networks.
This research is supported in part by the MASTODONS-2016 CNRS project 
APLIGOOGLE (see \url{http://www.quantware.ups-tlse.fr/APLIGOOGLE/}).
This work was granted access to the HPC resources of 
CALMIP (Toulouse) under the allocation 2016-P0110. 

%\vskip -1.5cm

\end{document}